\documentclass[%
 reprint,
superscriptaddress,
amssymb,
 aps,
]{revtex4-1}
\usepackage{amsmath}
\usepackage{graphicx}% Include figure files
\usepackage{dcolumn}% Align table columns on decimal point
\usepackage{bm}% bold math
\newcommand{\Vb}{V_{\rm{b}}}%					
\newcommand{\VTIBB}{\varphi_{\rm{BB}}}%		
\newcommand{\Vo}{W_0}%	
\newcommand{\dIdV}{\mathrm{d}I/\mathrm{d}V}%	

\begin{document}

%\preprint{APS/123-QED}

\title{
Poor electronic screening in lightly doped Mott insulators  \\observed with scanning tunneling microscopy
}% Force line breaks with \\

%thanks{A footnote to the article title}%

\author{I. Battisti}
\affiliation{Leiden Institute of Physics, Leiden University, Niels Bohrweg 2, 2333 CA Leiden, The Netherlands}
\author{V. Fedoseev}
\affiliation{Leiden Institute of Physics, Leiden University, Niels Bohrweg 2, 2333 CA Leiden, The Netherlands}
\author{K. M. Bastiaans}
\affiliation{Leiden Institute of Physics, Leiden University, Niels Bohrweg 2, 2333 CA Leiden, The Netherlands}
\author{A. de la Torre}
\affiliation{Institute for Quantum Information and Matter, California Institute of Technology, Pasadena, California 91125, USA}
\affiliation{Department of Physics, California Institute of Technology, Pasadena, California 91125, USA}
\author{R. S. Perry}
\affiliation{London Centre for Nanotechnology and UCL Centre for Materials Discovery, University College London, London WC1E 6BT, UK}
\author{F. Baumberger}
\affiliation{Department of Quantum Matter Physics, University of Geneva, 24 Quai Ernest-Ansermet, 1211 Geneva 4, Switzerland}
\affiliation{Swiss Light Source, Paul Scherrer Institute, CH-5232 Villigen PSI, Switzerland}
\author{M. P. Allan}
\affiliation{Leiden Institute of Physics, Leiden University, Niels Bohrweg 2, 2333 CA Leiden, The Netherlands}
%\affiliation{% Authors' institution and/or address}%

\date{\today}

\begin{abstract}

The effective Mott gap measured by scanning tunneling microscopy (STM) in the lightly doped Mott insulator $(\rm{Sr}_{1 -x}\rm{La}_x)_2\rm{IrO}_4$ differs greatly from values reported by photoemission and optical experiments. Here, we show that this is a consequence of the poor electronic screening of the tip-induced electric field in this material.  Such effects are well known from STM experiments on semiconductors, and go under the name of tip-induced band bending (TIBB). We show that this phenomenon also exists in the lightly doped Mott insulator  $(\rm{Sr}_{1 -x}\rm{La}_x)_2\rm{IrO}_4$ and that, at doping concentrations of $x\leq 4 \%$, it causes the measured energy gap in the sample density of states to be bigger than the one measured with other techniques. We develop a model able to retrieve the intrinsic energy gap leading to a value which is in rough agreement with other experiments, bridging the apparent contradiction. At doping $x \approx 5 \%$  we further observe circular features in the conductance layers that point to the emergence of a significant density of free carriers in this doping range, and to the presence of a small concentration of donor atoms. We illustrate the importance of considering the presence of TIBB when doing STM experiments on correlated-electron systems and discuss the similarities and differences between STM measurements on semiconductors and lightly doped Mott insulators.

%\begin{description}
%\end{description}
\end{abstract}

\pacs{Valid PACS appear here}% PACS, the Physics and Astronomy
\maketitle
\section{Introduction}
Mott insulators are a class of materials that should be metallic according to band theory, but are insulating due to strong electron-electron interactions.  When the Coulomb repulsion $U$ is much larger than the kinetic (hopping) energy $t$, the electrons localize on the atomic sites, resulting in the opening of a gap in the density of states between the so-called lower and upper Hubbard band. The chemical potential is located inside the gap and therefore the material is insulating. Quasi two-dimensional layered Mott insulators with perovskite crystal structure, with the cuprates being the most famous example, are of particular interest in condensed matter physics. In their parent state, without the insertion of extra carriers (doping), the localized spins arrange in an antiferromagnetic insulating ground state. But when lightly doped, these materials show a wide number of different behaviors. Famous examples are the coexistence of nanoscale stripes of metallic and insulating regions, charge ordering, the pseudogap phase and high-temperature superconductivity \cite{Keimer2015,Fujita2012,Kohsaka2012,Parker2010,Cai2015}. 

The physics of Mott insulators is radically different from the physics of semiconductors. In the latter, the gap around the chemical potential is a bandgap instead of a correlation-induced gap and the picture of independent and itinerant electrons is valid. However, in contrast to metals, Mott insulators and semiconductors share a reduced ability to screen electric fields. As a consequence, externally applied electric fields can partially penetrate the material. This can have important implications when performing STM experiments on such materials.
\begin{figure}
\centering
\includegraphics[width=8.5cm]{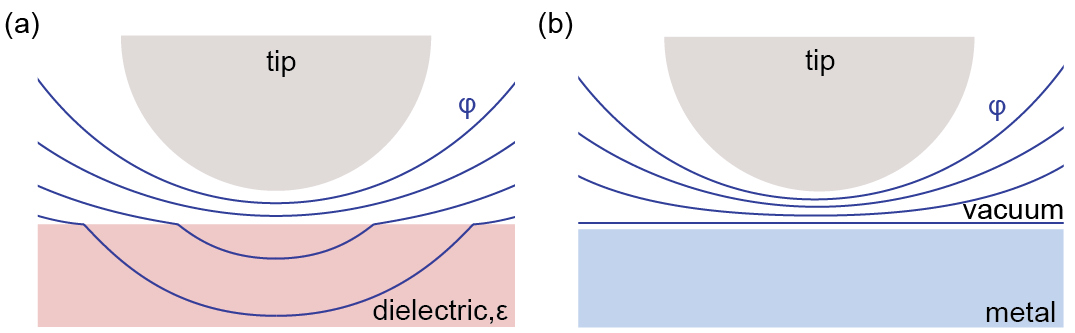}
\caption{Equipotential lines in STM experiments showing different screening of electric fields in different materials. (a), When STM experiments are performed on samples with a gapped density of states at the Fermi energy, the electric field can penetrate the sample due to the poor electronic screening. (b) In standard STM experiments on metals, the electric field generated by the tip is largely screened within the first atomic layer and there is no significant field penetration.}
\label{fig:fig1}
\end{figure}

Indeed, STM experiments on semiconductors  reveal that the electric field generated by the tip can partially penetrate the sample surface, causing an additional potential drop inside the sample [Fig. 1(a)]. Because the potential landscape changes in a way similar to how bands bend at semiconductor interfaces, this effect is known as tip-induced band bending (TIBB) \cite{Feenstra1987a, Feenstra2006, Weimer1989}. (Note how different this is from `conventional' STM experiments on metal surfaces with good electronic screening: When a metallic sample like copper is placed in the electric field  generated by the STM tip, the field is almost perfectly screened and there is no relevant field penetration [Fig. 1(b)].)

The phenomenon of TIBB has been widely studied in the semiconductor community and it can strongly affect the interpretation of STM data. For instance, the apparent gap measured with tunneling spectroscopy can significantly differ from the intrinsic bandgap in the density of states of the sample, as it has been observed e.g. on the surfaces of Ge(111) \cite{Feenstra2006a}, FeS$_2$(100) \cite{Herbert2013} and ZnO \cite{Sabitova2013}. Moreover, TIBB can also cause the ionization of donors/acceptors in the semiconductor \cite{Teichmann2008,Wijnheijmer2011, Loth2006}, and the effect has even been used in tip-induced quantum dot experiments \cite{Morgenstern2001}. Being able to quantitatively calculate TIBB is necessary for the interpretation of data: only if the values of TIBB are known, the intrinsic bandgap can be retrieved from the data, and the binding energies of the donors/acceptors can be extracted.
Using the known dielectric constants and carrier concentration, this is often done for semiconductors with a Poisson's equation solver developed by Feenstra \cite{feenstrasolver}, yielding apparent bandgaps around 15-20\% larger than the intrinsic ones \cite{Feenstra2006a,Herbert2013}.

TIBB is a direct consequence of poor electronic screening and therefore one might expect TIBB to be present not only in semiconductors, but in any other material with poor electronic screening. In fact, signatures of TIBB are observed for the lightly hole-doped oxychloride Ca$_2$CuO$_2$Cl$_2$ \cite{Kohsaka2012}, and poor electronic screening effects around charged impurities are observed for Fe dopants in the topological insulator Bi$_2$Se$_3$ \cite{Song2012}, for Co adatoms in graphene \cite{Brar2011} and possibly for  chiral defects in Sr$_3$Ir$_2$O$_7$ \cite{Okada2013}. TIBB has also been discussed for 2D transition metal dichalcogenides \cite{Ugeda2014} and for graphene systems \cite{Wong2015}. We expect that TIBB could affect measurements, or could even be used for gating, of topical materials with poor electronic screening, including iron-based superconductors, transition metal dichalcogenides, van der Waals heterostructures, or new topological materials.  However, other than in semiconductors and especially with respect to Mott insulators, the effects of TIBB have not been analyzed much.

In this article we concentrate  on the lightly electron-doped Mott insulator $(\rm{Sr}_{1 -x}\rm{La}_x)_2\rm{IrO}_4$, where we discover clear indications of electric field penetration inside the sample using  STM. We  develop a  model of electric field penetration in the absence of free carriers specifically for lightly doped Mott insulators where important material parameters are not known. This  allows us to calculate TIBB and to better understand the physics of the material, and to provide new insights for STM experiments on lightly doped Mott insulators in general. 

\section{\label{sec:ir}STM experiments on the iridate S\MakeLowercase{r}$_2$I\MakeLowercase{r}O$_4$}

The parent compound Sr$_2$IrO$_4$ belongs to the Ruddlesden-Popper series of perovskite iridates Sr$_{n+1}$Ir$_n$O$_{3n+1}$ with $n=1$. It is a quasi two-dimensional effective Mott insulator due to spin-orbit coupling enhanced correlations \cite{Rau2016}. Optical conductivity measurements on the parent compound lead to an extrapolated gap value of $\sim 400$~meV (with the first peak in the optical spectra positioned at $\sim 500$~meV) \cite{Kim2008, Moon2009}, which is in good agreement with theoretical calculations \cite{Kim2012}. Previous STM experiments investigated both undoped and doped Sr$_2$IrO$_4$ and Sr$_3$Ir$_2$O$_7$ (bilayer compound obtained for $n=2$). The bilayer Sr$_3$Ir$_2$O$_7$  has a smaller band gap, and this allowed STM measurements on the parent compound down to 4~K \cite{Okada2013}. When doping Sr$_3$Ir$_2$O$_7$ with electrons via La substitutions, first a nanoscale phase separation appears consisting of insulating regions and metallic puddles, and subsequently, for higher doping, a homogeneous metallic state emerges \cite{Hogan}.  The monolayer compound Sr$_2$IrO$_4$ with a nominal gap of $\sim 400$~meV \cite{Kim2008, Kim2012, Moon2009} is expected to be an insulator at low temperatures. Consequently, pioneering measurements are reported only at 77~K \cite{Dai2014, Nichols2014, Li2013} with mutually different gap values. Accidental doping was reported in one study \cite{Dai2014}; this is a possible cause  of the  different values. When $(\rm{Sr}_{1 -x}\rm{La}_x)_2\rm{IrO}_4$ is doped with electrons via La substitutions up to $x = 5\%$ (maximal doping level reported so far), measurements of the ab-plane resistivity show metallic behavior down to $\sim 50$~K followed by an upturn at lower temperature \cite{Chen2015,DelaTorre2015}. To what extent this behavior is intrinsic to $(\rm{Sr}_{1 -x}\rm{La}_x)_2\rm{IrO}_4$ is not understood. We know from previous STM measurements \cite{Battisti2017} that the doping concentration is not homogeneous within one sample, but can change on a length-scale of hundreds of micrometers, which might complicate the interpretation of transport data. Moreover, for $x\approx5\%$, samples show a nanoscale phase separation between Mott insulating regions with a gap of $\sim 500$~meV and pseudogap regions around clusters of La atoms \cite{Chen2015, Battisti2017}.

In this paper we present spectroscopic STM measurements on $(\rm{Sr}_{1 -x}\rm{La}_x)_2\rm{IrO}_4$. All data is measured below 8~K on atomically flat, SrO terminated surfaces, obtained by mechanical cleaving of the $(\rm{Sr}_{1 -x}\rm{La}_x)_2\rm{IrO}_4$ samples at T$ \sim 20$~K and $p=2\cdot 10^{-10}$~mbar. The STM topographs are taken in constant current mode, and the $\dIdV$ curves are measured using standard lock-in techniques with a modulation frequency $f =857$~Hz at constant tunneling conductance. We typically measure spectroscopic maps: For each pixel of a topograph we switch off the feedback and sweep the bias voltage $\Vb$ while measuring a spectrum of differential conductance $\dIdV$. This yields  a three-dimensional dataset that consists of a set of $\dIdV$ spectra measured on a fine 2-dimensional grid $(r_x,r_y)$, i.e.\ two spatial axis and one voltage (energy) axis. Mechanically ground PtIr tips are used for all measurements, with a tip radius of 20-30~nm estimated by scanning electron microscopy (SEM). The spectroscopic and topographic properties of the tips are tested on a crystalline Au(111) surface prepared \textit{in situ} by Ar ion sputtering and  annealing before measuring $(\rm{Sr}_{1 -x}\rm{La}_x)_2\rm{IrO}_4$.

\section{\label{sec:gap} A model for retrieving the intrinsic energy scales in the density of states }

\begin{figure}
\centering
\includegraphics[width=8.5cm]{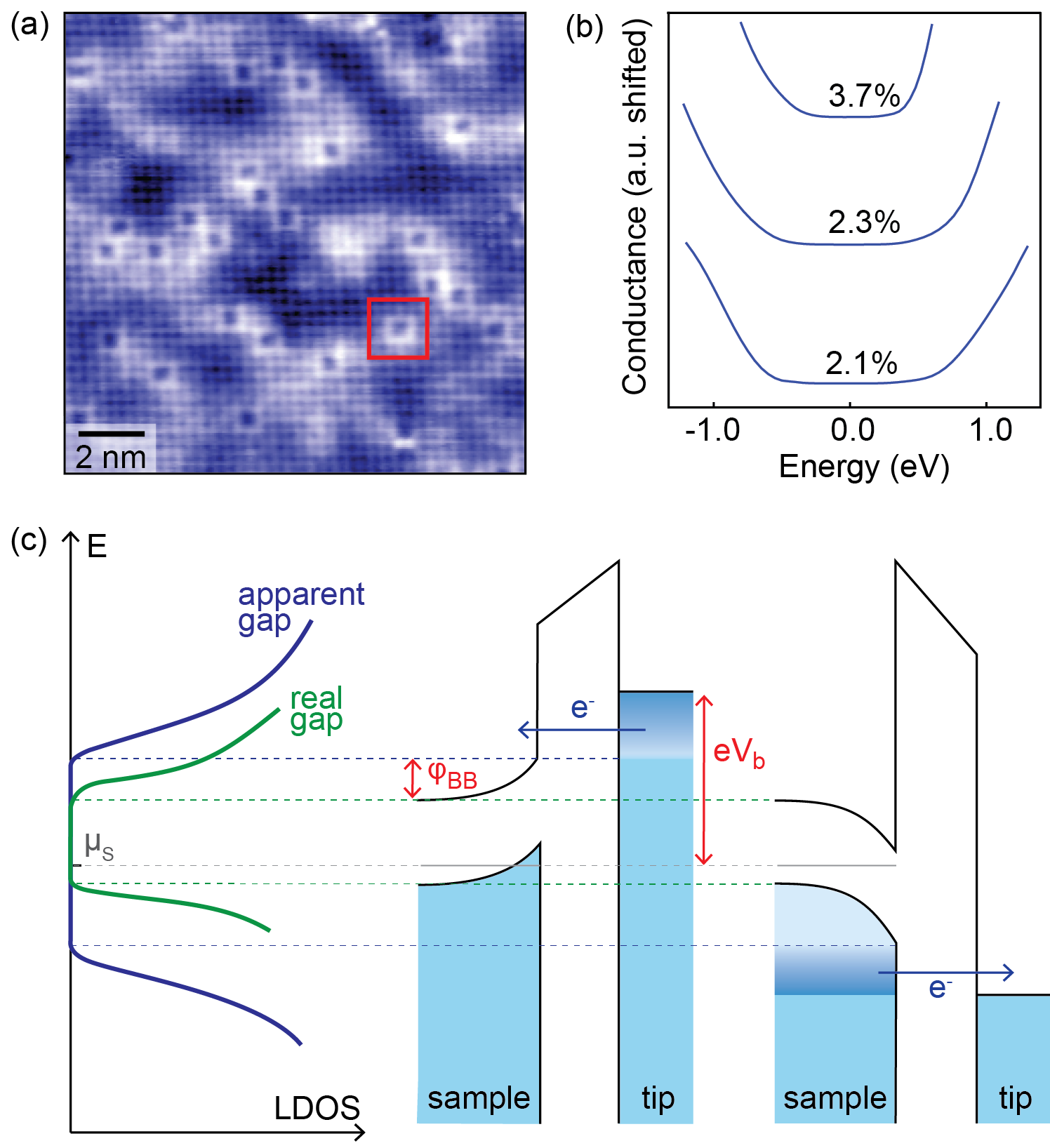}
\caption{(a) Topograph of a typical surface with $x=2\%$ doping. The setup conditions are $V_{\rm{s}}=-1.1  \,\rm{V}, I_{\rm{s}}=-200$~pA. La dopants (e.g. in the red square) are identified as dark atoms surrounded by a square of brighter conductance, as previously reported in Ref. \cite{Hogan}. (b) Raw tunneling $\dIdV$ spectra measured on $(\rm{Sr}_{1 -x}\rm{La}_x)_2\rm{IrO}_4$ samples. Each spectrum is the average of $10^4-10^5$ spectra measured in regions with different doping. (At zero doping, the samples are perfectly insulating prohibiting STM experiments even at 77~K.) (c) Qualitative explanation of TIBB illustrating why the measured gap exceeds the intrinsic gap. When a bias voltage $\Vb$ is applied between tip and sample, TIBB induces a voltage difference $\VTIBB$ between the bottom of the sample (grounded) and its surface.}
\label{fig:fig2}
\end{figure}

Figure 2(a) depicts a typical topograph for doping concentration $x\approx2\%$; the SrO lattice is visible with lattice constant $a_0\approx3.9$ \AA, and the white squares arise around the positions of La dopant atoms in the surface layer \cite{Hogan}. We previously found that up to a doping threshold of $x\approx4\%$ the material shows homogeneous insulating behaviour with an observed electronic gap of $\sim1$~eV. This is in disagreement with the values reported in literature from different techniques \cite{Kim2008,Moon2009,Dai2014, DelaTorre2015, Kim2012}, hinting towards the presence of TIBB. Figure 2(b) shows examples of these spectra: Each spectrum is the average of $10^4$ to $10^5$ spectra measured on a spectroscopic map in a field of view of $10\times10$~nm$^2$ at regions with different doping levels in the range $x\approx2-4\%$. Note that in the parent state,  $\rm{Sr}_2\rm{IrO}_4$ is insulating, yielding STM experiments impossible. Both at 4K and 77K, the tip  crashes during approach. We interpret this as evidence for the high quality of the sample and the absence of accidental doping in the parent state.

To show qualitatively how the observed apparent gap in the density of states (DOS) is affected by the presence of TIBB, let us consider a scanning tunneling spectroscopy experiment when TIBB is present [Fig.~2(c)]. When measuring a spectrum, we first set up the tip at ($V_{\rm{setup}}$, $I_{\rm{setup}}$) and go out of feedback. The bias voltage~$\Vb$ is then swept from positive to negative, while measuring the differential conductance $\dIdV(\Vb)$. For $\Vb>0$ the unoccupied states are probed, where electrons tunnel from the tip to the sample, while for $\Vb<0$ the occupied states are probed, where electrons tunnel from the sample to the tip. In the case of a gapped DOS as in a Mott insulator, the onset in the tunneling current occurs when the tip Fermi energy crosses the lower boundary of the upper Hubbard band or the upper boundary of the lower Hubbard band. Both events occur at higher absolute bias voltages~$\Vb$ in the presence of TIBB as the bands bend upwards for $\Vb>0$ and downwards for $\Vb<0$. Thus the apparent gap is wider than the real one when the tip electric field penetrates the sample.

In the following, we develop a model of electric field penetration in the absence of free carriers that allows us to calculate TIBB. We then calculate the tunneling current using Bardeen's tunneling equation amended to include TIBB. Thus, the model allows us to predict the measured differential conductance curves ($G= \dIdV$) in the presence of TIBB as a function of tip geometry, dielectric properties, tip-sample distance and difference in work function between the tip and the sample. We measure a series of spectra at different tip heights and we fix the parameters concerning the geometric and dielectric properties, remaining with only one free parameter that we can fit to the data. This allows us to extract the native density of states of the sample and to reconcile the results of our measurements with literature.

We consider a situation as depicted in Fig.~3(a). First, we need to find the band bending potential $\VTIBB$ at the point in the sample closest to the tip (point A). As a first approximation, we model the tip as a conductive charged sphere of radius $r$ at a distance $h$ from the sample, $h\ll r$, and the sample as a dielectric medium with dielectric constant $\varepsilon$ filling a half-space. We consider a bias voltage $\Vb$ applied between the tip and the bottom of the sample, which is grounded. 

Using the image charges method (see Appendix), we obtain the band bending potential  $\VTIBB$ at the point A in terms of the sphere potential $e\Vb$. The value of $\VTIBB$ obtained this way depends on the sphere radius $r$, the tip-sample distance $h$ and the dielectric constant $\varepsilon$. In the simplest approximation of a uniformly charged sphere, an analytic expression for the TIBB can be obtained:
\begin{equation}
\label{eq:TIBB1}
\VTIBB(\Vb, r,h,\varepsilon )=\frac{1}{1+\varepsilon \frac{h}{r}} \cdot (e\Vb - \Vo),
\end{equation}
where $\Vo$ represents the difference in work functions between the sample and the tip $\Vo=W_{\rm{sample}}-W_{\rm{tip}}$.
In the more realistic case of charge redistribution on the tip, a more general expression for TIBB needs to be considered, which we describe in the Appendix. Here, we absorb the proportionality in a new constant,
\begin{equation}
\label{eq:TIBB2}
\VTIBB(\Vb, r,h,\varepsilon )=F(r,h,\varepsilon) \cdot (e\Vb - \Vo).
\end{equation}

In order to calculate $\VTIBB$ for realistic parameters of our setup, we measure the typical tip radius for our tips as $r=25$~nm using SEM, and estimate the static dielectric constant of Sr$_2$IrO$_4$ as $\varepsilon=30$ for the parent compound (based on Ref.~\cite{Chikara2009a}). This value is a rough estimate, and we assume that it can still be applied in the case of the considered very low doping concentration.

For these parameters we find $F = 0.430$ for $h=0.3$~nm, $F=0.354$ for $h=0.5$~nm and $F=0.309$ for $h=0.7$~nm, setting for simplicity $\Vo=0$~eV. These analytic results agree within 1\% accuracy with finite element calculations obtained with the commercial software package COMSOL~\cite{comsol}.
We further use finite element simulations to compare our spherical tip approximation to the more realistic geometry of the tip, modeled as a metallic cone with the aperture of $20^{\circ}$ ending with an appropriate spherical segment. We find that with such a tip geometry, the value of $\VTIBB$ increases by 15-20\%, and conclude that our approximation of a spherical tip yields reliable results.

Next, we need to calculate $G = \dIdV$ in the presence of TIBB using Bardeen's tunneling equation \cite{Bardeen1961}, modified to include TIBB as described by Eq.~(\ref{eq:TIBB2}). 

The tunneling current in the presence of band bending at $T=0$~K, for a bias voltage $\Vb$ and tip-sample distance $h$, is given by
\begin{equation}
\label{eq:curr1}
\begin{split}
I(\Vb, h)& = \frac{4\pi e}{\hbar}\; |M(h)|^2 \\ 
& \times \!\int\displaylimits_0^{eV_{\rm{max}}}\!\!\mathrm{d} u\; g_s(\mu+u) \;g_t \big( \mu-eV_b+\VTIBB(\Vb, h)+u \big),
\end{split}
\end{equation}
with $eV_{\rm{max}}=e\Vb-\VTIBB(\Vb, h)$, where $\VTIBB$ is taken from Eq.~(\ref{eq:TIBB2}). The sample and tip DOS are respectively $g_s$ and $g_t$, and $|M(h)|^2$ represents the tunneling matrix elements.
In the assumption of constant tip DOS $g_t$, the tunneling current simplifies to
\begin{equation}
\label{eq:curr2}
I(\Vb, h)=\frac{4\pi e}{\hbar}|M(h)|^2 \ g_t\int\displaylimits_0^{eV_{\rm{max}}} \mathrm{d}u \; g_s(\mu+u).
\end{equation}
We can calculate the differential conductance $G(\Vb,h)=\partial I(\Vb,h)/\partial\Vb$ simply by taking the derivative of Eq. (\ref{eq:curr2}) with respect to $\Vb$, obtaining
\begin{equation}
\label{eq:cond1}
\begin{split}
G(\Vb,h)&=\frac{4\pi e^2}{\hbar}\bigg(1-\frac{\partial \VTIBB(\Vb, h)}{\partial \Vb}\bigg)|M(h)|^2 \\
& \times g_t \;g_s\big(\mu+e\Vb-\VTIBB(\Vb, h)\big).
\end{split}
\end{equation}

Figure 3(b) shows a series of $G\equiv \dIdV$ spectra measured subsequently at the same location with increasing tip-sample distances on a sample with 2.2~\% doping. A clear dependence on the setup conditions is visible. The setup bias voltage is kept constant at $V_{\rm{s}}=1.5$ V and the setup current $I_{\rm{s}}$ ranges from 600 pA to 10 pA, covering almost two orders of magnitude.

\begin{figure}
\centering
\includegraphics[width=8.5cm]{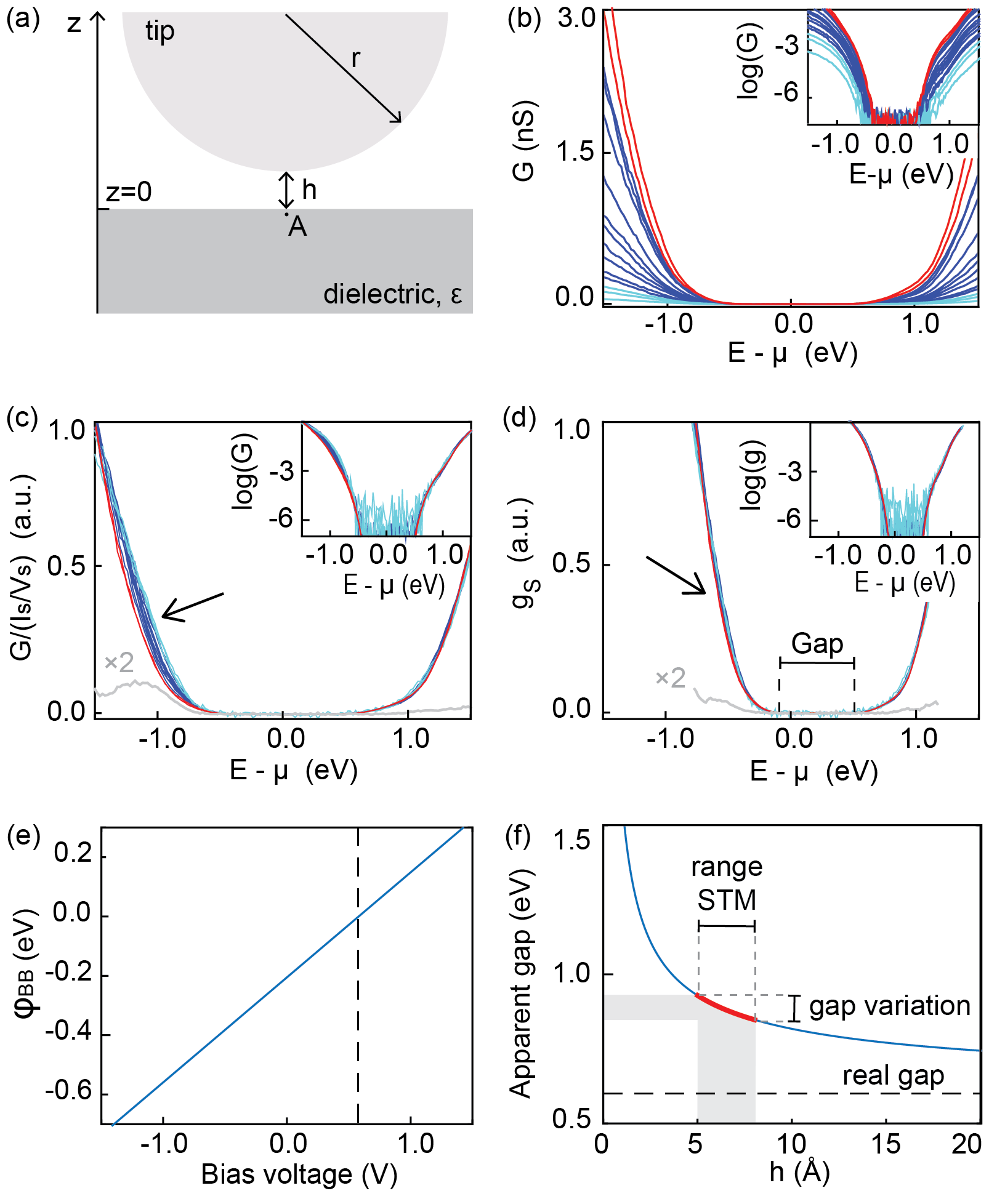}
\caption{(a) Tip-sample geometry ($r$ and $h$ are not to scale) as used for the electric potential calculation. (b) $G\equiv dI/dV$ spectra measured at different tip-sample distances $h$ on a sample with 2.2~\% doping. The bias setup voltage $V_{s}$ is fixed to 1.5~V and the current $I_{s}$ goes from 600~pA (light blue) to 10~pA (red). In the inset the same plot is shown on a logarithmic scale. (c) The same spectra as in panel (b), each normalized by its its setup junction resistance $I_{s}/V_{s}$. The gray line shows the standard deviation $\sigma(G)$ calculated for each energy, multiplied by a factor two.  The inset shows the same plot with logarithmic scale. (d) Extracted intrinsic DOS $g_s$ as a function of $E$ obtained from Eq. (\ref{eq:end}), after minimization of the model parameter $W_0$, yielding an  intrinsic gap of 600~meV. The gray line shows the standard deviation $\sigma(g_s)$ calculated for each energy, multiplied by a factor two. Since the rescaling of the curves causes different horizontal axes for each curve, we calculate $\sigma(g_s)$ over extrapolated values of $g_s$ at equally spaced energies. The inset shows the same plot with logarithmic scale. (e) Calculated  $\VTIBB(V_b, h=5\textrm{\AA})$ function for different bias voltages. The dashed line represents the voltage corresponding to the work function difference $W_0$, at which no band bending is present. (f) Calculated apparent gap width in the sample DOS for different tip-sample distances. Within distances where typical STM experiments are performed the variation is relatively small compared to the difference with the real gap value.}
\label{fig:fig3}
\end{figure}

The differences between the spectra are due to the different tip-sample distances $h$, which are mainly included in the unknown tunneling matrix elements $|M(h)|^2$. Following Ref. \cite{Feenstra1987}, we eliminate $|M(h)|^2$ by normalizing the differential conductance $G(\Vb,h)$ by the setup current divided
by the voltage:
\begin{equation}
\label{eq:cond2}
\bar{G}(\Vb,h)\equiv \frac{G(\Vb,h)}{I_{\rm{s}}/V_{\rm{s}}}
\end{equation}
In absence of TIBB, $\bar{G}$ becomes independent of $h$, and such normalized spectra should collapse on a single curve. 

We apply Eq.~(\ref{eq:cond2}) to the data in Fig.~3(b), plotting the result in Fig.~3(c). It is immediately noticed that the curves do not collapse exactly on each other, the biggest differences arising for negative energies (see arrow). We quantify this difference by the standard deviations calculated for each energy [shown as the gray line in Fig. 3(c)]. We attribute these differences in the normalized spectra to the presence of TIBB and thus further modeling is required to extract the intrinsic sample DOS.

To do so, we calculate an effective bias voltage $V^{\rm{eff}}(h)$ for each tip-sample distance $h$ such that 
\begin{equation}
\label{eq:transf}
eV_{\rm{s}}-\VTIBB(V_{\rm{s}}, h) \equiv eV^{\rm{eff}}(h)-\VTIBB(V^{\rm{eff}}(h),h_0)
\end{equation}
for a fixed tip-sample distance $h_0$.

Using Eq. (\ref{eq:transf}), we rewrite Baarden's tunneling equation as:
\begin{equation}
\label{eq:int}
\int\limits_0^{eV_{\rm{s}}-\VTIBB(V_{\rm{s}},h)} g_s(\mu+u)du=\frac{I(V^{\rm{eff}}(h), h_0)}{\frac{4\pi e}{\hbar}|M(h_0)|^2 g_t}.
\end{equation}
By inserting Eq.~(\ref{eq:int}) into Eq.~(\ref{eq:cond1}) divided by the setup conditions, we can extract the intrinsic density of states $g_s(\mu +u)$ from measured $G(h)$ curves at different heights:
\begin{multline}
\label{eq:end}
g_s(\mu+u)=\frac{G(h)}{I_{\rm{s}}/V_{\rm{s}}}\frac{1}{1-\frac{\partial \VTIBB(\Vb,h)}{\partial \Vb}}
\frac{I(V^{\rm{eff}}(h), h_0)}{\frac{4\pi e^2}{\hbar}|M(h_0)|^2g_t},
\end{multline}
where $u=e\Vb-\VTIBB(\Vb, h)$. 

The parameters present in the model are ${\varepsilon}$, $r$, the difference in work functions $W_0$, the minimal tip-sample distance $h_{\rm{min}}$ and the exponential prefactor $\kappa$ of the tunneling current $I=I_0 \cdot e^{-\kappa h}$. 
We keep $r$ and $\varepsilon$ fixed at the values mentioned previously. 
We estimate $h_{\rm{min}}=5\,\textrm{\AA}$ as a typical tunneling distance for 1~G$\Omega$ tunneling resistance for this material. From measured $I(z)$ curves, we determine $\kappa=1.1$~\AA$^{-1}$. Thus the only free parameter in Eq. (9) is $W_0$.

We apply our model to the data of Fig.~3(b), extracting the parameter $W_0$ as the value that minimizes the error function $\Omega=\int[\sigma (g_s)]^2 $, where $\sigma(g_s)$ are the standard deviations of the $g_s$ curves for each energy. Minimization gives a work function difference between the tip and the sample of $W_0=0.55$~eV.

We show the result of the application of our model to the data in Fig.~3(d). It can immediately be noticed that the spectra  are rescaled in energy, leading to a gap value of 600~meV. This value is in good agreement with literature \cite{Kim2008, Moon2009, Kim2012}, allowing us to reconcile our measurement to the other techniques. Moreover, a comparison of Fig.~3(c) and 3(d), clearly shows that the curves overlap better after the procedure, as quantified by comparing the standard deviations (gray lines in both panels).

Further, we show in Fig.~3(e) the calculated value of TIBB for the point on the sample surface closest to the tip, as a function of the applied bias voltage. Note that for bias voltage corresponding to $W_0$, there is no TIBB (flat band condition). 
Figure 3(f) shows the dimension of the apparent gap as a function of the tip-sample distance $h$. While there is a remarkable difference between the intrinsic gap value and the apparent gap, we want to stress that, within the values of $h$ in which STM experiments are conducted, the variation of the apparent gap is relatively small. Therefore, even if measurements do not show sizable dependence on setup conditions, TIBB might be present, and further analysis might be required to retrieve the intrinsic energy scales.

In spite of the simplicity of the model, we are able to capture qualitatively the behaviour of the system at low doping levels. Two important conclusions can be drawn from this analysis. First, the restored energy scale for $g_s$ allows us to reconcile the onset of the lower Hubbard band in STM and ARPES. The onset of the lower Hubbard band at $\sim-0.1$~eV obtained when our model is applied to the data agrees well with the photoemission value reported for samples at low-doping levels in Ref.~\cite{DelaTorre2015, Wang2013}, and the full gap of 0.6~eV agrees roughly with the optical data from Ref.~\cite{Kim2008, Moon2009}. Second, TIBB can strongly affect the measured DOS. Consequently, when measuring samples with poor electronic screening, the eventuality of field penetration must be taken into account, and further analysis might be required to retrieve the native density of states.

We note that this model is not applicable on samples in the higher doping level of 5\%, where we observe pseudogap puddles emerging in the density of states. As described in Section \ref{sec:bubbles}, we still find evidence of field penetration in samples with 5\% doping, however our model is not able to detect the effects of TIBB at this doping level. We presume that the model is not effective because for higher doping levels the number of free carriers can no longer be neglected, breaking our first assumption. 

\section{\label{sec:bubbles}Bubbles in the conductance layers}

\begin{figure}
\centering
\includegraphics[width=8.5cm]{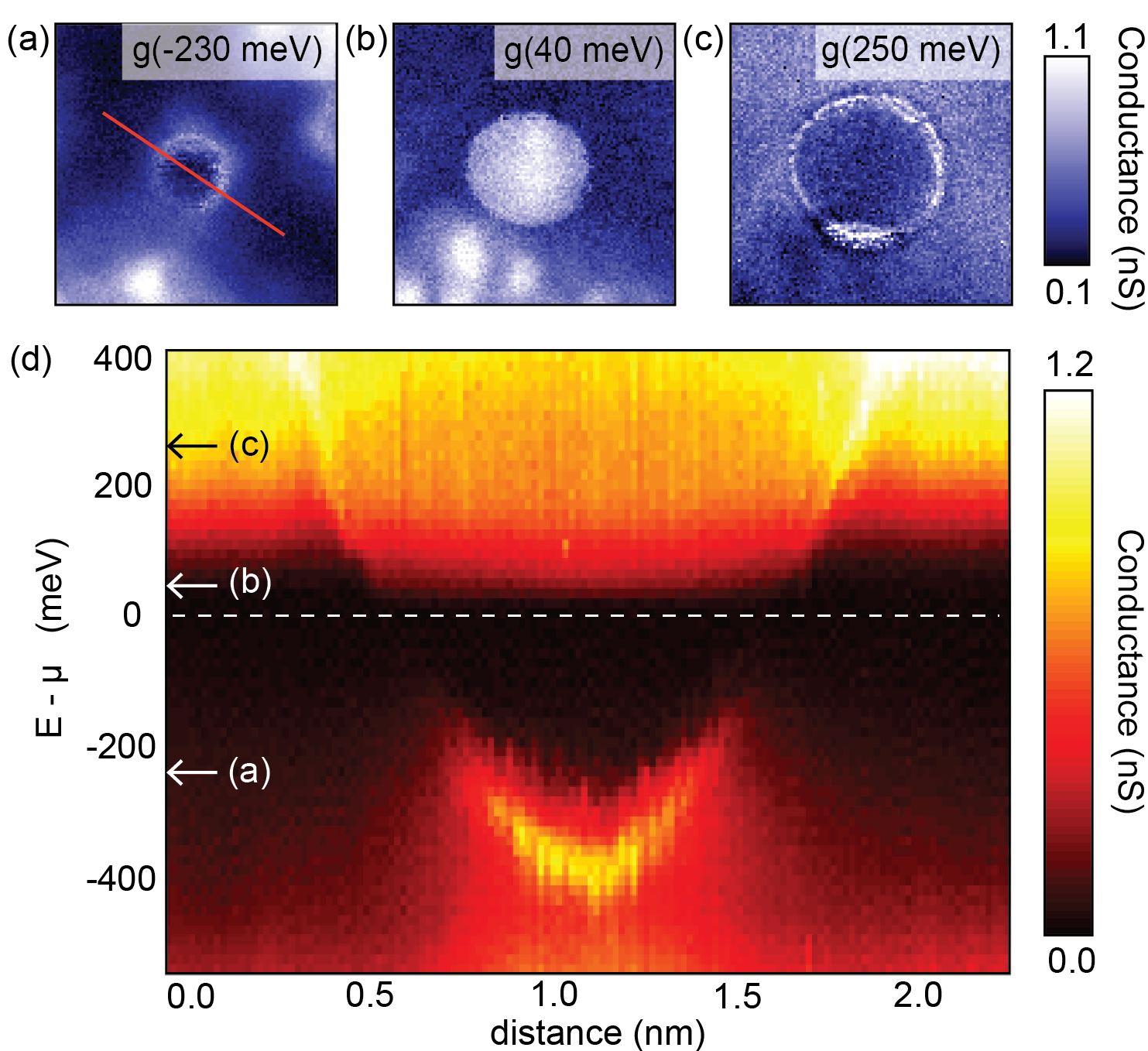}
\caption{Visualization of a tip-induced band bending bubble in $(\rm{Sr}_{1 -x}\rm{La}_x)_2\rm{IrO}_4$ at $x\approx5$\%. (a,b,c) Conductance layers of a field of view of $3\times3\,\rm{nm}^2$ at g(-230~meV), g(40~meV), g(250~meV). (d) Cross section in energy [($E,r$) plot] of the bubble along the red line in (a). The hyperbolic profile is due to the increasing diameter of the bubble with increasing energy. The arrows indicate the energies at which the conductance layers shown in panels (a,b,c) are extracted.}
\label{fig:fig4}
\end{figure}

In the samples with higher doping levels ($x \approx 5 \%$), we observe a different signature of field penetration: Circular rings of enhanced conductance appear in the  layers of constant energy of the conductance maps. In the following, we will refer to these features as `bubbles'. Their diameter increases with energy, as shown in Fig.~\ref{fig:fig4}(a-c), causing hyperbolas of enhanced conductance in a cross section in energy [($E,r$)~plot], as it can be seen in Fig.~\ref{fig:fig4}(d). We shall see that the bubbles are generated by the presence of a low concentration of specific impurity atoms which can be used as a probe to better understand the field penetration in the material. We will come back to the nature of these impurity atoms later in this section.

Very similar features have been observed in semiconductors, where they are identified as markers of ionization/empty state filling of donors or acceptors induced by the vicinity of the STM tip. `Bubbles' in semiconductors have been thoroughly studied because they can help in extracting material parameters such as the donors' binding energy. This was done for instance for Si donors in GaAs \cite{Teichmann2008,Wijnheijmer2011}, for which it was further demonstrated that donors closer to the surface have an enhanced binding energy with respect to the bulk \cite{Wijnheijmer2009}. Effects of charge manipulation by the STM tip and enhanced binding energy closer to the surface were also reported for Mn acceptors in InAs and GaAs \cite{Marczinowski2008, Lee2011} and for donors in ZnO \cite{Zheng2012, Zheng2013}. Moreover, bubbles due to TIBB effects have also been reported when using a scanning capacitance probe to image transport in two-dimensional electron gas in AlGaAs/GaAs heterostructures \cite{Steele2005}.

We note that signatures of finite field penetration resembling the bubbles observed in our samples are also found in other correlated-electron systems, such as the lightly hole-doped oxychloride Ca$_2$CuO$_2$Cl$_2$ \cite{Kohsaka2012} and possibly the correlated iridates Sr$_3$Ir$_2$O$_7$ and $\rm{Sr}_3(\rm{Ir}_{1  -x}\rm{Ru}_x)_2\rm{O}_7$ \cite{Okada2013, Walkup2016}. However, these bubbles have never been discussed in details for a correlated-electron system.

We expect that the mechanisms that lead to the formation of the bubbles in our samples are the same as in semiconductors, and we refer to Ref.~\cite{Teichmann2008, Wijnheijmer2011} for a detailed description of the processes. 

Here we emphasize that the impurity atoms in our samples are identified as electron donors, that each of these donors generates one hyperbola as in Fig.~4(d), and that the two parts of the hyperbola lying above and below the chemical potential come from two different tunneling processes. For $\Vb>0$, the enhanced conductance is due to the ionization of the donor, which locally changes the potential landscape in the sample. In this process, the electrons tunnel from the tip to the bulk of the sample, therefore the bubble becomes visible only after the onset of the upper Hubbard band.  
For $V_b<0$, the enhanced conductance is instead caused by the opening of an additional tunneling channel. In this process, electrons tunnel from the sample bulk to the tip via the donor state. The bubble's diameter in this part of the hyperbola reflects the extension of the donor wave function in real space. 
Both processes are triggered at a specific value of $\VTIBB$, causing the hyperbola to follow a constant $\VTIBB$ contour. We emphasize that the two parts of the hyperbola will lie on the same constant $\VTIBB$ contour only when the sample chemical potential roughly coincides with the onset of the upper Hubbard band, otherwise they might be shifted in energy.

In a typical spectroscopic map, we can usually identify several bubbles which start to emerge at different energies. Figure~5(a) shows the topograph of a field of view of 17$\times$17 nm$^2$ with doping level of 5.5\%, where we count 180 dopant atoms on the surface. In the same field of view, the conductance layers show the appearance of only $\sim15$ bubbles [Fig. 5(b)]. In general, the number of bubbles that we observe corresponds to $\lesssim 10\%$ of the total number of La dopants present on the surface. We can therefore exclude that La dopants in their normal state cause the appearance of the bubbles. Our best hypothesis on the nature of the bubbles is that they originate either from some special chemical state of the La atoms (for instance an oxygen vacancy next to the La atom) or from Pt atoms that substitute for the Ir atoms. The latter could originate from the Pt crucible where the samples were grown. 

It is important to note that the bubbles are not influenced by and do not influence the phase-separated density of state of the sample. In Fig.~5(c) we show a conductance layer with a black contour indicating the border between the phase-separated pseudogap and Mott states \cite{Battisti2017}. As it can be noticed, the bubbles originate from both Mott regions (where there are no dopant atoms) and pseudogap regions (forming around agglomerates of dopant atoms), and when they cross the sharp border between two regions their shape is not affected. Moreover, the phase-separated landscape and the emerging order that we describe in Ref.~\cite{Battisti2017} are not influenced by the presence of the bubbles. 

\begin{figure}[]
\centering
\includegraphics[width=8.5cm]{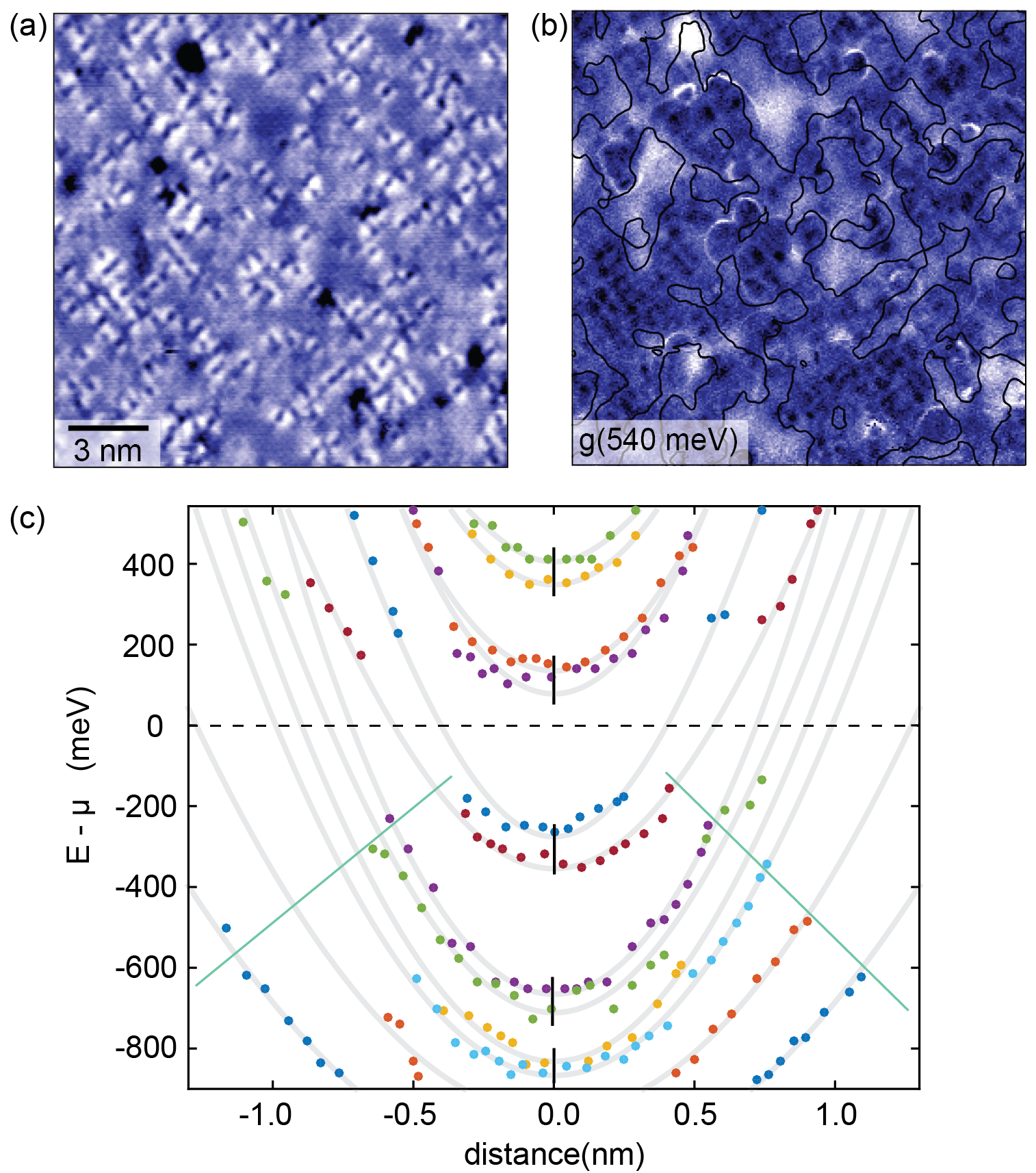}
\caption{(a) Topograph of a sample with $\sim5\%$ doping in a field of view of 17$\times$17 nm$^2$. The setup conditions are ($V_{\rm{s}}=460\, \rm{meV}, I_{\rm{s}}=300\, \rm{pA}$). We count 180 La dopants, easily identified on the surface as bright squares. (b) Conductance layer at g(540 meV) in the same field of view. We observe $\sim15$ circular bubbles of different sizes. The black line indicates the border of the phase separation between the pseudogap puddles (around clusters of dopant atoms) and the Mott areas \cite{Battisti2017}. (c) Hyperbolas extracted from all the bubbles appearing in the dataset shown in (a),(b). The gray lines are fits to the hyperbolas, added as guide to the eye. Most of the hyperbolas are only above or only below the sample chemical potential, but for a few of them both the positive and the negative energy parts are visible. The two green straight lines emphasize the increasing maximal bubbles' diameter with increasing donor depth below the surface. The vertical black lines indicate the grouping of hyperbolas starting at similar threshold potentials.}
\label{fig:fig5}
\end{figure}

Unfortunately, the model that  we developed for the low doping level samples is unable to grasp the physics of the samples with doping $x\approx5\%$, due to the presence of free carriers in the latter case.  We can still make some important qualitative observations by plotting in Fig. 5(c) all the hyperbolas extracted from the bubbles in Fig. 5(b):

(i), The bubbles start to appear at different threshold potentials. The threshold potential is an indication of the donor depth below the surface \cite{Wijnheijmer2009}, with donors that lie deeper below the surface having a lower threshold potential. We therefore conclude that we observe bubbles originating form donors located at different depths. 

(ii), For the lower part of the hyperbola, the maximum bubble's diameter gets smaller for donors closer to the surface. Since the maximum diameter reflects the real space extension of the donor wave function, this gives evidence for enhanced binding energy for donors closer to the surface \cite{Wijnheijmer2009}. 

(iii), Most of the bubbles can be grouped as starting at roughly the same threshold potential (within an error of 50 meV), therefore probably originating from donors at the same depth below the surface, i.e. belonging to the same crystal layer. In Fig.~5(c) this is indicated by the short vertical black lines.

Before concluding, we would like to emphasize a last important point that might tell us something more about the material. The typical lateral extension of the bubbles in our samples ranges from 1 to 2~nm. This is significantly lower than in semiconductors where, for example, the typical extension of bubbles due to Si donors in GaAs is 10~nm. Among the factors that can influence the extension of the bubbles are the tip radius, the concentration of free carriers and the material's electrical permittivity. We can exclude that the tip radius is the cause for the small extension of the bubbles, as one would need to have an unrealistically small tip radius to reproduce the bubbles.
It is known from transport studies that the resistivity in $(\rm{Sr}_{1 -x}\rm{La}_x)_2\rm{IrO}_4$ is lower in the ab-crystal plane than along the c-axis \cite{Chikara2009a}, although with diminishing strength upon doping \cite{Ge2011}. Moreover, the electrical permittivity of Sr$_2$IrO$_4$ is anisotropic \cite{Chikara2009a}, and we believe that this could also be a factor influencing the shape of the hyperbolas. We can only speculate that the small extension of the bubbles is related to these effects; in any case it is evidence for the strongly anisotropic electronic structure of the material.

\section{Summary}
Performing STM measurements on materials with poor electronic screening can lead to tip-induced band bending (TIBB) and an apparent energy gap in the sample DOS that is larger than the intrinsic value. TIBB and its consequences for STM measurements are well known for semiconductors and its hallmarks have been observed (but not yet discussed) for correlated electrons systems \cite{Kohsaka2012, Walkup2016}. 

Here, we report detailed measurements of TIBB in the iridate $(\rm{Sr}_{1 -x}\rm{La}_x)_2\rm{IrO}_4$. We show that TIBB has to be taken into account when measuring samples near the insulating state: Even if energy scales seem to vary only little with set-up conditions, there might still be large differences with the intrinsic energy scales.

We present a simple model that, under the assumption of the absence of free carriers, allows us to calculate TIBB and to reconstruct the intrinsic energy scales from the measured $\dIdV$ spectra, reconciling STM measurements with ARPES and optics \cite{DelaTorre2015,Kim2008}. We furthermore show the limits of this model at higher doping levels, where we observe a different consequence of TIBB appearing as bubbles in the conductance layers. These bubbles indicate the emergence of qualitatively different electric field screening between the ab-plane and the c-axis.

\begin{acknowledgments}
We thank J. Jobst, P. M. Koenraad, M. Morgenstern, J. van Ruitenbeek, and J. Zaanen for valuable discussions. We acknowledge funding from the Netherlands Organization for Scientific Research (NWO/OCW) as part of the Frontiers of Nanoscience (NanoFront) programme and the VIDI talent scheme (Project No. 680-47-536).
\end{acknowledgments}

%%%%%%%%
\appendix*
\section{Image charges method for a charged sphere in front of a dielectric sample}

In order to calculate the electric potential $\VTIBB$, we make use of the image charges method to solve the configuration of a charged-metallic sphere of radius $r$ kept at distance $h$ in front of a dielectric sample with dielectric constant $\varepsilon$ [Fig. 3(a)]. The method of image charges consists in finding an equivalent set of point charges that satisfies the same boundary conditions of the original geometry at the dielectric surface and the sphere surface \cite{Jackson1999}.

For our configuration, the solution is given by an infinite series of image charges with diminishing absolute value. This set of image charges is built in the following recurrent sequence: A charge $q$ is added to an uncharged sphere, which is the equivalent of a point charge in the centre of the sphere $(q; r+h)$. This point charge induces a image charge in the dielectric medium $(-kq; -(r+h))$, where $k=\frac{\varepsilon-1}{\varepsilon+1}$, which in turn induces a dipole image on the sphere $(\frac{-kqr}{2(r+h)}; r+h)$ and $(\frac{+kqr}{2(r+h)}, r+h-\frac{r^2}{2(r+h)})$, and so on. 

The electric potential in the whole space is then given by
\begin{equation}
\varphi(\bm{r})=\frac{\kappa}{4\pi\varepsilon_0}\sum\frac{q_i}{|\bm{r}-\bm{r}_i|},
\end{equation}

where for $z\geq0$, $\kappa=1$ and $(q_i, \bm{r}_i)$ are the initial charge and all the image charges induced on the sphere and in the sample; for $z<0$, $\kappa=\frac{2}{1+\varepsilon}$ and $(q_i, \bm{r}_i)$ are the initial charge and all the image charges induced on the sphere \cite{Jackson1999}.

To calculate $\VTIBB$, we compute the electrical potential on the tip surface and at the point of the sample surface closest to the tip. We extract the parameter $F(\varepsilon,r,h)$ (Section \ref{sec:gap}) as the ratio of these two potentials.

Note that the simplified situation of a uniformly charged sphere that can be replaced with a single point charge at the center of the sphere underestimates $\VTIBB$ by a factor of two for our setup $(r\sim25\, \textrm{nm}, h\sim0.5\,\textrm{nm}, \varepsilon\sim30)$. Consequently, it is important to take the full charge redistribution into account.

\bibliography{ms}

\begin{thebibliography}{46}%
\makeatletter
\providecommand \@ifxundefined [1]{%
 \@ifx{#1\undefined}
}%
\providecommand \@ifnum [1]{%
 \ifnum #1\expandafter \@firstoftwo
 \else \expandafter \@secondoftwo
 \fi
}%
\providecommand \@ifx [1]{%
 \ifx #1\expandafter \@firstoftwo
 \else \expandafter \@secondoftwo
 \fi
}%
\providecommand \natexlab [1]{#1}%
\providecommand \enquote  [1]{``#1''}%
\providecommand \bibnamefont  [1]{#1}%
\providecommand \bibfnamefont [1]{#1}%
\providecommand \citenamefont [1]{#1}%
\providecommand \href@noop [0]{\@secondoftwo}%
\providecommand \href [0]{\begingroup \@sanitize@url \@href}%
\providecommand \@href[1]{\@@startlink{#1}\@@href}%
\providecommand \@@href[1]{\endgroup#1\@@endlink}%
\providecommand \@sanitize@url [0]{\catcode `\\12\catcode `\$12\catcode
  `\&12\catcode `\#12\catcode `\^12\catcode `\_12\catcode `\%12\relax}%
\providecommand \@@startlink[1]{}%
\providecommand \@@endlink[0]{}%
\providecommand \url  [0]{\begingroup\@sanitize@url \@url }%
\providecommand \@url [1]{\endgroup\@href {#1}{\urlprefix }}%
\providecommand \urlprefix  [0]{URL }%
\providecommand \Eprint [0]{\href }%
\providecommand \doibase [0]{http://dx.doi.org/}%
\providecommand \selectlanguage [0]{\@gobble}%
\providecommand \bibinfo  [0]{\@secondoftwo}%
\providecommand \bibfield  [0]{\@secondoftwo}%
\providecommand \translation [1]{[#1]}%
\providecommand \BibitemOpen [0]{}%
\providecommand \bibitemStop [0]{}%
\providecommand \bibitemNoStop [0]{.\EOS\space}%
\providecommand \EOS [0]{\spacefactor3000\relax}%
\providecommand \BibitemShut  [1]{\csname bibitem#1\endcsname}%
\let\auto@bib@innerbib\@empty
%</preamble>
\bibitem [{\citenamefont {Keimer}\ \emph {et~al.}(2015)\citenamefont {Keimer},
  \citenamefont {Kivelson}, \citenamefont {Norman}, \citenamefont {Uchida},\
  and\ \citenamefont {Zaanen}}]{Keimer2015}%
  \BibitemOpen
  \bibfield  {author} {\bibinfo {author} {\bibfnamefont {B.}~\bibnamefont
  {Keimer}}, \bibinfo {author} {\bibfnamefont {S.~A.}\ \bibnamefont
  {Kivelson}}, \bibinfo {author} {\bibfnamefont {M.~R.}\ \bibnamefont
  {Norman}}, \bibinfo {author} {\bibfnamefont {S.}~\bibnamefont {Uchida}}, \
  and\ \bibinfo {author} {\bibfnamefont {J.}~\bibnamefont {Zaanen}},\
  }\href@noop {} {\bibfield  {journal} {\bibinfo  {journal} {Nature}\ }\textbf
  {\bibinfo {volume} {518}},\ \bibinfo {pages} {179} (\bibinfo {year}
  {2015})}\BibitemShut {NoStop}%
\bibitem [{\citenamefont {Fujita}(2012)}]{Fujita2012}%
  \BibitemOpen
  \bibfield  {author} {\bibinfo {author} {\bibfnamefont {K.}~\bibnamefont
  {Fujita}},\ }in\ \href@noop {} {\emph {\bibinfo {booktitle} {Theor. Methods
  Strongly Correl. Syst.}}}\ (\bibinfo  {publisher} {Springer Berlin
  Heidelberg},\ \bibinfo {year} {2012})\ Chap.~\bibinfo {chapter}
  {3}\BibitemShut {NoStop}%
\bibitem [{\citenamefont {Kohsaka}\ \emph {et~al.}(2012)\citenamefont
  {Kohsaka}, \citenamefont {Hanaguri}, \citenamefont {Azuma}, \citenamefont
  {Takano}, \citenamefont {Davis},\ and\ \citenamefont {Takagi}}]{Kohsaka2012}%
  \BibitemOpen
  \bibfield  {author} {\bibinfo {author} {\bibfnamefont {Y.}~\bibnamefont
  {Kohsaka}}, \bibinfo {author} {\bibfnamefont {T.}~\bibnamefont {Hanaguri}},
  \bibinfo {author} {\bibfnamefont {M.}~\bibnamefont {Azuma}}, \bibinfo
  {author} {\bibfnamefont {M.}~\bibnamefont {Takano}}, \bibinfo {author}
  {\bibfnamefont {J.~C.}\ \bibnamefont {Davis}}, \ and\ \bibinfo {author}
  {\bibfnamefont {H.}~\bibnamefont {Takagi}},\ }\href@noop {} {\bibfield
  {journal} {\bibinfo  {journal} {Nat. Phys.}\ }\textbf {\bibinfo {volume}
  {8}},\ \bibinfo {pages} {534} (\bibinfo {year} {2012})}\BibitemShut {NoStop}%
\bibitem [{\citenamefont {Parker}\ \emph {et~al.}(2010)\citenamefont {Parker},
  \citenamefont {Aynajian}, \citenamefont {da~Silva~Neto}, \citenamefont
  {Pushp}, \citenamefont {Ono}, \citenamefont {Wen}, \citenamefont {Xu},
  \citenamefont {Gu},\ and\ \citenamefont {Yazdani}}]{Parker2010}%
  \BibitemOpen
  \bibfield  {author} {\bibinfo {author} {\bibfnamefont {C.~V.}\ \bibnamefont
  {Parker}}, \bibinfo {author} {\bibfnamefont {P.}~\bibnamefont {Aynajian}},
  \bibinfo {author} {\bibfnamefont {E.~H.}\ \bibnamefont {da~Silva~Neto}},
  \bibinfo {author} {\bibfnamefont {A.}~\bibnamefont {Pushp}}, \bibinfo
  {author} {\bibfnamefont {S.}~\bibnamefont {Ono}}, \bibinfo {author}
  {\bibfnamefont {J.}~\bibnamefont {Wen}}, \bibinfo {author} {\bibfnamefont
  {Z.}~\bibnamefont {Xu}}, \bibinfo {author} {\bibfnamefont {G.}~\bibnamefont
  {Gu}}, \ and\ \bibinfo {author} {\bibfnamefont {A.}~\bibnamefont {Yazdani}},\
  }\href@noop {} {\bibfield  {journal} {\bibinfo  {journal} {Nature}\ }\textbf
  {\bibinfo {volume} {468}},\ \bibinfo {pages} {677} (\bibinfo {year}
  {2010})}\BibitemShut {NoStop}%
\bibitem [{\citenamefont {Cai}\ \emph {et~al.}(2016)\citenamefont {Cai},
  \citenamefont {Ruan}, \citenamefont {Peng}, \citenamefont {Ye}, \citenamefont
  {Li}, \citenamefont {Hao}, \citenamefont {Zhou}, \citenamefont {Lee},\ and\
  \citenamefont {Wang}}]{Cai2015}%
  \BibitemOpen
  \bibfield  {author} {\bibinfo {author} {\bibfnamefont {P.}~\bibnamefont
  {Cai}}, \bibinfo {author} {\bibfnamefont {W.}~\bibnamefont {Ruan}}, \bibinfo
  {author} {\bibfnamefont {Y.}~\bibnamefont {Peng}}, \bibinfo {author}
  {\bibfnamefont {C.}~\bibnamefont {Ye}}, \bibinfo {author} {\bibfnamefont
  {X.}~\bibnamefont {Li}}, \bibinfo {author} {\bibfnamefont {Z.}~\bibnamefont
  {Hao}}, \bibinfo {author} {\bibfnamefont {X.}~\bibnamefont {Zhou}}, \bibinfo
  {author} {\bibfnamefont {D.-H.}\ \bibnamefont {Lee}}, \ and\ \bibinfo
  {author} {\bibfnamefont {Y.}~\bibnamefont {Wang}},\ }\href@noop {} {\bibfield
   {journal} {\bibinfo  {journal} {Nat. Phys.}\ }\textbf {\bibinfo {volume}
  {12}},\ \bibinfo {pages} {1047} (\bibinfo {year} {2016})}\BibitemShut
  {NoStop}%
\bibitem [{\citenamefont {Feenstra}\ and\ \citenamefont
  {Stroscio}(1987)}]{Feenstra1987a}%
  \BibitemOpen
  \bibfield  {author} {\bibinfo {author} {\bibfnamefont {R.~M.}\ \bibnamefont
  {Feenstra}}\ and\ \bibinfo {author} {\bibfnamefont {J.~a.}\ \bibnamefont
  {Stroscio}},\ }\href@noop {} {\bibfield  {journal} {\bibinfo  {journal} {J.
  Vac. Sci. Technol. B}\ }\textbf {\bibinfo {volume} {5}},\ \bibinfo {pages}
  {295} (\bibinfo {year} {1987})}\BibitemShut {NoStop}%
\bibitem [{\citenamefont {Feenstra}\ \emph
  {et~al.}(2006{\natexlab{a}})\citenamefont {Feenstra}, \citenamefont {Dong},
  \citenamefont {Semtsiv},\ and\ \citenamefont {Masselink}}]{Feenstra2006}%
  \BibitemOpen
  \bibfield  {author} {\bibinfo {author} {\bibfnamefont {R.~M.}\ \bibnamefont
  {Feenstra}}, \bibinfo {author} {\bibfnamefont {Y.}~\bibnamefont {Dong}},
  \bibinfo {author} {\bibfnamefont {M.~P.}\ \bibnamefont {Semtsiv}}, \ and\
  \bibinfo {author} {\bibfnamefont {W.~T.}\ \bibnamefont {Masselink}},\
  }\href@noop {} {\bibfield  {journal} {\bibinfo  {journal} {Nanotechnology}\
  }\textbf {\bibinfo {volume} {18}},\ \bibinfo {pages} {044015} (\bibinfo
  {year} {2006}{\natexlab{a}})}\BibitemShut {NoStop}%
\bibitem [{\citenamefont {Weimer}\ \emph {et~al.}(1989)\citenamefont {Weimer},
  \citenamefont {Kramar},\ and\ \citenamefont {Baldeschwieler}}]{Weimer1989}%
  \BibitemOpen
  \bibfield  {author} {\bibinfo {author} {\bibfnamefont {M.}~\bibnamefont
  {Weimer}}, \bibinfo {author} {\bibfnamefont {J.}~\bibnamefont {Kramar}}, \
  and\ \bibinfo {author} {\bibfnamefont {J.~D.}\ \bibnamefont
  {Baldeschwieler}},\ }\href@noop {} {\bibfield  {journal} {\bibinfo  {journal}
  {Phys. Rev. B}\ }\textbf {\bibinfo {volume} {39}},\ \bibinfo {pages} {5572}
  (\bibinfo {year} {1989})}\BibitemShut {NoStop}%
\bibitem [{\citenamefont {Feenstra}\ \emph
  {et~al.}(2006{\natexlab{b}})\citenamefont {Feenstra}, \citenamefont {Lee},
  \citenamefont {Kang}, \citenamefont {Meyer},\ and\ \citenamefont
  {Rieder}}]{Feenstra2006a}%
  \BibitemOpen
  \bibfield  {author} {\bibinfo {author} {\bibfnamefont {R.~M.}\ \bibnamefont
  {Feenstra}}, \bibinfo {author} {\bibfnamefont {J.~Y.}\ \bibnamefont {Lee}},
  \bibinfo {author} {\bibfnamefont {M.~H.}\ \bibnamefont {Kang}}, \bibinfo
  {author} {\bibfnamefont {G.}~\bibnamefont {Meyer}}, \ and\ \bibinfo {author}
  {\bibfnamefont {K.~H.}\ \bibnamefont {Rieder}},\ }\href
  {http://link.aps.org/doi/10.1103/PhysRevB.73.035310} {\bibfield  {journal}
  {\bibinfo  {journal} {Phys. Rev. B}\ }\textbf {\bibinfo {volume} {73}},\
  \bibinfo {pages} {035310} (\bibinfo {year} {2006}{\natexlab{b}})}\BibitemShut
  {NoStop}%
\bibitem [{\citenamefont {Herbert}\ \emph {et~al.}(2013)\citenamefont
  {Herbert}, \citenamefont {Krishnamoorthy}, \citenamefont {{Van Vliet}},\ and\
  \citenamefont {Yildiz}}]{Herbert2013}%
  \BibitemOpen
  \bibfield  {author} {\bibinfo {author} {\bibfnamefont {F.~W.}\ \bibnamefont
  {Herbert}}, \bibinfo {author} {\bibfnamefont {A.}~\bibnamefont
  {Krishnamoorthy}}, \bibinfo {author} {\bibfnamefont {K.~J.}\ \bibnamefont
  {{Van Vliet}}}, \ and\ \bibinfo {author} {\bibfnamefont {B.}~\bibnamefont
  {Yildiz}},\ }\href@noop {} {\bibfield  {journal} {\bibinfo  {journal} {Surf.
  Sci.}\ }\textbf {\bibinfo {volume} {618}},\ \bibinfo {pages} {53} (\bibinfo
  {year} {2013})}\BibitemShut {NoStop}%
\bibitem [{\citenamefont {Sabitova}\ \emph {et~al.}(2013)\citenamefont
  {Sabitova}, \citenamefont {Ebert}, \citenamefont {Lenz}, \citenamefont
  {Schaafhausen}, \citenamefont {Ivanova}, \citenamefont {D{\"{a}}hne},
  \citenamefont {Hoffmann}, \citenamefont {Dunin-Borkowski}, \citenamefont
  {F{\"{o}}rster}, \citenamefont {Grandidier},\ and\ \citenamefont
  {Eisele}}]{Sabitova2013}%
  \BibitemOpen
  \bibfield  {author} {\bibinfo {author} {\bibfnamefont {A.}~\bibnamefont
  {Sabitova}}, \bibinfo {author} {\bibfnamefont {P.}~\bibnamefont {Ebert}},
  \bibinfo {author} {\bibfnamefont {A.}~\bibnamefont {Lenz}}, \bibinfo {author}
  {\bibfnamefont {S.}~\bibnamefont {Schaafhausen}}, \bibinfo {author}
  {\bibfnamefont {L.}~\bibnamefont {Ivanova}}, \bibinfo {author} {\bibfnamefont
  {M.}~\bibnamefont {D{\"{a}}hne}}, \bibinfo {author} {\bibfnamefont
  {A.}~\bibnamefont {Hoffmann}}, \bibinfo {author} {\bibfnamefont {R.~E.}\
  \bibnamefont {Dunin-Borkowski}}, \bibinfo {author} {\bibfnamefont
  {A.}~\bibnamefont {F{\"{o}}rster}}, \bibinfo {author} {\bibfnamefont
  {B.}~\bibnamefont {Grandidier}}, \ and\ \bibinfo {author} {\bibfnamefont
  {H.}~\bibnamefont {Eisele}},\ }\href@noop {} {\bibfield  {journal} {\bibinfo
  {journal} {Appl. Phys. Lett.}\ }\textbf {\bibinfo {volume} {102}},\ \bibinfo
  {pages} {021608} (\bibinfo {year} {2013})}\BibitemShut {NoStop}%
\bibitem [{\citenamefont {Teichmann}\ \emph {et~al.}(2008)\citenamefont
  {Teichmann}, \citenamefont {Wenderoth}, \citenamefont {Loth}, \citenamefont
  {Ulbrich}, \citenamefont {Garleff}, \citenamefont {Wijnheijmer},\ and\
  \citenamefont {Koenraad}}]{Teichmann2008}%
  \BibitemOpen
  \bibfield  {author} {\bibinfo {author} {\bibfnamefont {K.}~\bibnamefont
  {Teichmann}}, \bibinfo {author} {\bibfnamefont {M.}~\bibnamefont
  {Wenderoth}}, \bibinfo {author} {\bibfnamefont {S.}~\bibnamefont {Loth}},
  \bibinfo {author} {\bibfnamefont {R.~G.}\ \bibnamefont {Ulbrich}}, \bibinfo
  {author} {\bibfnamefont {J.~K.}\ \bibnamefont {Garleff}}, \bibinfo {author}
  {\bibfnamefont {A.~P.}\ \bibnamefont {Wijnheijmer}}, \ and\ \bibinfo {author}
  {\bibfnamefont {P.~M.}\ \bibnamefont {Koenraad}},\ }\href@noop {} {\bibfield
  {journal} {\bibinfo  {journal} {Phys. Rev. Lett.}\ }\textbf {\bibinfo
  {volume} {101}},\ \bibinfo {pages} {076103} (\bibinfo {year}
  {2008})}\BibitemShut {NoStop}%
\bibitem [{\citenamefont {Wijnheijmer}\ \emph {et~al.}(2011)\citenamefont
  {Wijnheijmer}, \citenamefont {Garleff}, \citenamefont {Teichmann},
  \citenamefont {Wenderoth}, \citenamefont {Loth},\ and\ \citenamefont
  {Koenraad}}]{Wijnheijmer2011}%
  \BibitemOpen
  \bibfield  {author} {\bibinfo {author} {\bibfnamefont {A.~P.}\ \bibnamefont
  {Wijnheijmer}}, \bibinfo {author} {\bibfnamefont {J.~K.}\ \bibnamefont
  {Garleff}}, \bibinfo {author} {\bibfnamefont {K.}~\bibnamefont {Teichmann}},
  \bibinfo {author} {\bibfnamefont {M.}~\bibnamefont {Wenderoth}}, \bibinfo
  {author} {\bibfnamefont {S.}~\bibnamefont {Loth}}, \ and\ \bibinfo {author}
  {\bibfnamefont {P.~M.}\ \bibnamefont {Koenraad}},\ }\href@noop {} {\bibfield
  {journal} {\bibinfo  {journal} {Phys. Rev. B}\ }\textbf {\bibinfo {volume}
  {84}},\ \bibinfo {pages} {125310} (\bibinfo {year} {2011})}\BibitemShut
  {NoStop}%
\bibitem [{\citenamefont {Loth}\ \emph {et~al.}(2006)\citenamefont {Loth},
  \citenamefont {Wenderoth}, \citenamefont {Winking}, \citenamefont {Ulbrich},
  \citenamefont {Malzer},\ and\ \citenamefont {D{\"{o}}hler}}]{Loth2006}%
  \BibitemOpen
  \bibfield  {author} {\bibinfo {author} {\bibfnamefont {S.}~\bibnamefont
  {Loth}}, \bibinfo {author} {\bibfnamefont {M.}~\bibnamefont {Wenderoth}},
  \bibinfo {author} {\bibfnamefont {L.}~\bibnamefont {Winking}}, \bibinfo
  {author} {\bibfnamefont {R.~G.}\ \bibnamefont {Ulbrich}}, \bibinfo {author}
  {\bibfnamefont {S.}~\bibnamefont {Malzer}}, \ and\ \bibinfo {author}
  {\bibfnamefont {G.~H.}\ \bibnamefont {D{\"{o}}hler}},\ }\href@noop {}
  {\bibfield  {journal} {\bibinfo  {journal} {Phys. Rev. Lett.}\ }\textbf
  {\bibinfo {volume} {96}},\ \bibinfo {pages} {066403} (\bibinfo {year}
  {2006})}\BibitemShut {NoStop}%
\bibitem [{\citenamefont {Morgenstern}\ \emph {et~al.}(2001)\citenamefont
  {Morgenstern}, \citenamefont {Gudmundsson}, \citenamefont {Dombrowski},
  \citenamefont {Wittneven},\ and\ \citenamefont
  {Wiesendanger}}]{Morgenstern2001}%
  \BibitemOpen
  \bibfield  {author} {\bibinfo {author} {\bibfnamefont {M.}~\bibnamefont
  {Morgenstern}}, \bibinfo {author} {\bibfnamefont {V.}~\bibnamefont
  {Gudmundsson}}, \bibinfo {author} {\bibfnamefont {R.}~\bibnamefont
  {Dombrowski}}, \bibinfo {author} {\bibfnamefont {C.}~\bibnamefont
  {Wittneven}}, \ and\ \bibinfo {author} {\bibfnamefont {R.}~\bibnamefont
  {Wiesendanger}},\ }\href@noop {} {\bibfield  {journal} {\bibinfo  {journal}
  {Phys. Rev. B}\ }\textbf {\bibinfo {volume} {63}},\ \bibinfo {pages}
  {201301(R)} (\bibinfo {year} {2001})}\BibitemShut {NoStop}%
\bibitem [{\citenamefont {Feenstra}()}]{feenstrasolver}%
  \BibitemOpen
  \bibfield  {author} {\bibinfo {author} {\bibfnamefont {R.~M.}\ \bibnamefont
  {Feenstra}},\ }\href@noop {} {}\bibinfo {howpublished} {{SEMITIP, version 6
  (2001)}, http://www.andrew.cmu.edu/user/feenstra/semitip\_v6}\BibitemShut
  {NoStop}%
\bibitem [{\citenamefont {Song}\ \emph {et~al.}(2012)\citenamefont {Song},
  \citenamefont {Jiang}, \citenamefont {Wang}, \citenamefont {Li},
  \citenamefont {Wang}, \citenamefont {He}, \citenamefont {Chen}, \citenamefont
  {Ma},\ and\ \citenamefont {Xue}}]{Song2012}%
  \BibitemOpen
  \bibfield  {author} {\bibinfo {author} {\bibfnamefont {C.~L.}\ \bibnamefont
  {Song}}, \bibinfo {author} {\bibfnamefont {Y.~P.}\ \bibnamefont {Jiang}},
  \bibinfo {author} {\bibfnamefont {Y.~L.}\ \bibnamefont {Wang}}, \bibinfo
  {author} {\bibfnamefont {Z.}~\bibnamefont {Li}}, \bibinfo {author}
  {\bibfnamefont {L.}~\bibnamefont {Wang}}, \bibinfo {author} {\bibfnamefont
  {K.}~\bibnamefont {He}}, \bibinfo {author} {\bibfnamefont {X.}~\bibnamefont
  {Chen}}, \bibinfo {author} {\bibfnamefont {X.~C.}\ \bibnamefont {Ma}}, \ and\
  \bibinfo {author} {\bibfnamefont {Q.~K.}\ \bibnamefont {Xue}},\ }\href@noop
  {} {\bibfield  {journal} {\bibinfo  {journal} {Phys. Rev. B}\ }\textbf
  {\bibinfo {volume} {86}},\ \bibinfo {pages} {045441} (\bibinfo {year}
  {2012})}\BibitemShut {NoStop}%
\bibitem [{\citenamefont {Brar}\ \emph {et~al.}(2011)\citenamefont {Brar},
  \citenamefont {Decker}, \citenamefont {Solowan}, \citenamefont {Wang},
  \citenamefont {Maserati}, \citenamefont {Chan}, \citenamefont {Lee},
  \citenamefont {Girit}, \citenamefont {Zettl}, \citenamefont {Louie},
  \citenamefont {Cohen},\ and\ \citenamefont {Crommie}}]{Brar2011}%
  \BibitemOpen
  \bibfield  {author} {\bibinfo {author} {\bibfnamefont {V.~W.}\ \bibnamefont
  {Brar}}, \bibinfo {author} {\bibfnamefont {R.}~\bibnamefont {Decker}},
  \bibinfo {author} {\bibfnamefont {H.-M.}\ \bibnamefont {Solowan}}, \bibinfo
  {author} {\bibfnamefont {Y.}~\bibnamefont {Wang}}, \bibinfo {author}
  {\bibfnamefont {L.}~\bibnamefont {Maserati}}, \bibinfo {author}
  {\bibfnamefont {K.~T.}\ \bibnamefont {Chan}}, \bibinfo {author}
  {\bibfnamefont {H.}~\bibnamefont {Lee}}, \bibinfo {author} {\bibfnamefont
  {{\c{C}}.~O.}\ \bibnamefont {Girit}}, \bibinfo {author} {\bibfnamefont
  {A.}~\bibnamefont {Zettl}}, \bibinfo {author} {\bibfnamefont {S.~G.}\
  \bibnamefont {Louie}}, \bibinfo {author} {\bibfnamefont {M.~L.}\ \bibnamefont
  {Cohen}}, \ and\ \bibinfo {author} {\bibfnamefont {M.~F.}\ \bibnamefont
  {Crommie}},\ }\href@noop {} {\bibfield  {journal} {\bibinfo  {journal} {Nat.
  Phys.}\ }\textbf {\bibinfo {volume} {7}},\ \bibinfo {pages} {43} (\bibinfo
  {year} {2011})}\BibitemShut {NoStop}%
\bibitem [{\citenamefont {Okada}\ \emph {et~al.}(2013)\citenamefont {Okada},
  \citenamefont {Walkup}, \citenamefont {Lin}, \citenamefont {Dhital},
  \citenamefont {Chang}, \citenamefont {Khadka}, \citenamefont {Zhou},
  \citenamefont {Jeng}, \citenamefont {Paranjape}, \citenamefont {Bansil},
  \citenamefont {Wang}, \citenamefont {Wilson},\ and\ \citenamefont
  {Madhavan}}]{Okada2013}%
  \BibitemOpen
  \bibfield  {author} {\bibinfo {author} {\bibfnamefont {Y.}~\bibnamefont
  {Okada}}, \bibinfo {author} {\bibfnamefont {D.}~\bibnamefont {Walkup}},
  \bibinfo {author} {\bibfnamefont {H.}~\bibnamefont {Lin}}, \bibinfo {author}
  {\bibfnamefont {C.}~\bibnamefont {Dhital}}, \bibinfo {author} {\bibfnamefont
  {T.-R.}\ \bibnamefont {Chang}}, \bibinfo {author} {\bibfnamefont
  {S.}~\bibnamefont {Khadka}}, \bibinfo {author} {\bibfnamefont
  {W.}~\bibnamefont {Zhou}}, \bibinfo {author} {\bibfnamefont {H.-T.}\
  \bibnamefont {Jeng}}, \bibinfo {author} {\bibfnamefont {M.}~\bibnamefont
  {Paranjape}}, \bibinfo {author} {\bibfnamefont {A.}~\bibnamefont {Bansil}},
  \bibinfo {author} {\bibfnamefont {Z.}~\bibnamefont {Wang}}, \bibinfo {author}
  {\bibfnamefont {S.~D.}\ \bibnamefont {Wilson}}, \ and\ \bibinfo {author}
  {\bibfnamefont {V.}~\bibnamefont {Madhavan}},\ }\href@noop {} {\bibfield
  {journal} {\bibinfo  {journal} {Nat. Mater.}\ }\textbf {\bibinfo {volume}
  {12}},\ \bibinfo {pages} {707} (\bibinfo {year} {2013})}\BibitemShut
  {NoStop}%
\bibitem [{\citenamefont {Ugeda}\ \emph {et~al.}(2014)\citenamefont {Ugeda},
  \citenamefont {Bradley}, \citenamefont {Shi}, \citenamefont {Jornada},
  \citenamefont {Zhang}, \citenamefont {Qiu}, \citenamefont {Ruan},
  \citenamefont {Mo}, \citenamefont {Hussain}, \citenamefont {Shen},
  \citenamefont {Wang}, \citenamefont {Louie},\ and\ \citenamefont
  {Crommie}}]{Ugeda2014}%
  \BibitemOpen
  \bibfield  {author} {\bibinfo {author} {\bibfnamefont {M.~M.}\ \bibnamefont
  {Ugeda}}, \bibinfo {author} {\bibfnamefont {A.~J.}\ \bibnamefont {Bradley}},
  \bibinfo {author} {\bibfnamefont {S.-f.}\ \bibnamefont {Shi}}, \bibinfo
  {author} {\bibfnamefont {F.~H.}\ \bibnamefont {Jornada}}, \bibinfo {author}
  {\bibfnamefont {Y.}~\bibnamefont {Zhang}}, \bibinfo {author} {\bibfnamefont
  {D.~Y.}\ \bibnamefont {Qiu}}, \bibinfo {author} {\bibfnamefont
  {W.}~\bibnamefont {Ruan}}, \bibinfo {author} {\bibfnamefont {S.-K.}\
  \bibnamefont {Mo}}, \bibinfo {author} {\bibfnamefont {Z.}~\bibnamefont
  {Hussain}}, \bibinfo {author} {\bibfnamefont {Z.-x.}\ \bibnamefont {Shen}},
  \bibinfo {author} {\bibfnamefont {F.}~\bibnamefont {Wang}}, \bibinfo {author}
  {\bibfnamefont {S.~G.}\ \bibnamefont {Louie}}, \ and\ \bibinfo {author}
  {\bibfnamefont {M.~F.}\ \bibnamefont {Crommie}},\ }\href@noop {} {\bibfield
  {journal} {\bibinfo  {journal} {Nat. Mat.}\ }\textbf {\bibinfo {volume}
  {13}},\ \bibinfo {pages} {1091} (\bibinfo {year} {2014})}\BibitemShut
  {NoStop}%
\bibitem [{\citenamefont {Wong}\ \emph {et~al.}(2015)\citenamefont {Wong},
  \citenamefont {Wang}, \citenamefont {Jung}, \citenamefont {Tsai},
  \citenamefont {Jung}, \citenamefont {Tollabimazraehno}, \citenamefont
  {Rasool}, \citenamefont {Watanabe}, \citenamefont {Taniguchi}, \citenamefont
  {Zettl}, \citenamefont {Adam}, \citenamefont {Macdonald},\ and\ \citenamefont
  {Crommie}}]{Wong2015}%
  \BibitemOpen
  \bibfield  {author} {\bibinfo {author} {\bibfnamefont {D.}~\bibnamefont
  {Wong}}, \bibinfo {author} {\bibfnamefont {Y.}~\bibnamefont {Wang}}, \bibinfo
  {author} {\bibfnamefont {J.}~\bibnamefont {Jung}}, \bibinfo {author}
  {\bibfnamefont {H.-Z.}\ \bibnamefont {Tsai}}, \bibinfo {author}
  {\bibfnamefont {H.~S.}\ \bibnamefont {Jung}}, \bibinfo {author}
  {\bibfnamefont {S.}~\bibnamefont {Tollabimazraehno}}, \bibinfo {author}
  {\bibfnamefont {H.}~\bibnamefont {Rasool}}, \bibinfo {author} {\bibfnamefont
  {K.}~\bibnamefont {Watanabe}}, \bibinfo {author} {\bibfnamefont
  {T.}~\bibnamefont {Taniguchi}}, \bibinfo {author} {\bibfnamefont
  {A.}~\bibnamefont {Zettl}}, \bibinfo {author} {\bibfnamefont
  {S.}~\bibnamefont {Adam}}, \bibinfo {author} {\bibfnamefont {A.~H.}\
  \bibnamefont {Macdonald}}, \ and\ \bibinfo {author} {\bibfnamefont {M.~F.}\
  \bibnamefont {Crommie}},\ }\href@noop {} {\bibfield  {journal} {\bibinfo
  {journal} {Phys. Rev. B}\ }\textbf {\bibinfo {volume} {92}},\ \bibinfo
  {pages} {155409} (\bibinfo {year} {2015})}\BibitemShut {NoStop}%
\bibitem [{\citenamefont {Rau}\ \emph {et~al.}(2016)\citenamefont {Rau},
  \citenamefont {Lee},\ and\ \citenamefont {Lee}}]{Rau2016}%
  \BibitemOpen
  \bibfield  {author} {\bibinfo {author} {\bibfnamefont {J.~G.}\ \bibnamefont
  {Rau}}, \bibinfo {author} {\bibfnamefont {E.~K.}\ \bibnamefont {Lee}}, \ and\
  \bibinfo {author} {\bibfnamefont {H.~Y.}\ \bibnamefont {Lee}},\ }\href@noop
  {} {\bibfield  {journal} {\bibinfo  {journal} {Ann. Rev. Condens. Matter
  Phys.}\ }\textbf {\bibinfo {volume} {7}},\ \bibinfo {pages} {195} (\bibinfo
  {year} {2016})}\BibitemShut {NoStop}%
\bibitem [{\citenamefont {Kim}\ \emph {et~al.}(2008)\citenamefont {Kim},
  \citenamefont {Jin}, \citenamefont {Moon}, \citenamefont {Kim}, \citenamefont
  {Park}, \citenamefont {Leem}, \citenamefont {Yu}, \citenamefont {Noh},
  \citenamefont {Kim}, \citenamefont {Oh}, \citenamefont {Park}, \citenamefont
  {Durairaj}, \citenamefont {Cao},\ and\ \citenamefont {Rotenberg}}]{Kim2008}%
  \BibitemOpen
  \bibfield  {author} {\bibinfo {author} {\bibfnamefont {B.~J.}\ \bibnamefont
  {Kim}}, \bibinfo {author} {\bibfnamefont {H.}~\bibnamefont {Jin}}, \bibinfo
  {author} {\bibfnamefont {S.~J.}\ \bibnamefont {Moon}}, \bibinfo {author}
  {\bibfnamefont {J.~Y.}\ \bibnamefont {Kim}}, \bibinfo {author} {\bibfnamefont
  {B.~G.}\ \bibnamefont {Park}}, \bibinfo {author} {\bibfnamefont {C.~S.}\
  \bibnamefont {Leem}}, \bibinfo {author} {\bibfnamefont {J.}~\bibnamefont
  {Yu}}, \bibinfo {author} {\bibfnamefont {T.~W.}\ \bibnamefont {Noh}},
  \bibinfo {author} {\bibfnamefont {C.}~\bibnamefont {Kim}}, \bibinfo {author}
  {\bibfnamefont {S.~J.}\ \bibnamefont {Oh}}, \bibinfo {author} {\bibfnamefont
  {J.~H.}\ \bibnamefont {Park}}, \bibinfo {author} {\bibfnamefont
  {V.}~\bibnamefont {Durairaj}}, \bibinfo {author} {\bibfnamefont
  {G.}~\bibnamefont {Cao}}, \ and\ \bibinfo {author} {\bibfnamefont
  {E.}~\bibnamefont {Rotenberg}},\ }\href@noop {} {\bibfield  {journal}
  {\bibinfo  {journal} {Phys. Rev. Lett.}\ }\textbf {\bibinfo {volume} {101}},\
  \bibinfo {pages} {076402} (\bibinfo {year} {2008})}\BibitemShut {NoStop}%
\bibitem [{\citenamefont {Moon}\ \emph {et~al.}(2009)\citenamefont {Moon},
  \citenamefont {Jin}, \citenamefont {Choi}, \citenamefont {Lee}, \citenamefont
  {Seo}, \citenamefont {Yu}, \citenamefont {Cao}, \citenamefont {Noh},\ and\
  \citenamefont {Lee}}]{Moon2009}%
  \BibitemOpen
  \bibfield  {author} {\bibinfo {author} {\bibfnamefont {S.~J.}\ \bibnamefont
  {Moon}}, \bibinfo {author} {\bibfnamefont {H.}~\bibnamefont {Jin}}, \bibinfo
  {author} {\bibfnamefont {W.~S.}\ \bibnamefont {Choi}}, \bibinfo {author}
  {\bibfnamefont {J.~S.}\ \bibnamefont {Lee}}, \bibinfo {author} {\bibfnamefont
  {S.~S.~A.}\ \bibnamefont {Seo}}, \bibinfo {author} {\bibfnamefont
  {J.}~\bibnamefont {Yu}}, \bibinfo {author} {\bibfnamefont {G.}~\bibnamefont
  {Cao}}, \bibinfo {author} {\bibfnamefont {T.~W.}\ \bibnamefont {Noh}}, \ and\
  \bibinfo {author} {\bibfnamefont {Y.~S.}\ \bibnamefont {Lee}},\ }\href@noop
  {} {\bibfield  {journal} {\bibinfo  {journal} {Phys. Rev. B}\ }\textbf
  {\bibinfo {volume} {80}},\ \bibinfo {pages} {195110} (\bibinfo {year}
  {2009})}\BibitemShut {NoStop}%
\bibitem [{\citenamefont {Kim}\ \emph {et~al.}(2012)\citenamefont {Kim},
  \citenamefont {Khaliullin},\ and\ \citenamefont {Min}}]{Kim2012}%
  \BibitemOpen
  \bibfield  {author} {\bibinfo {author} {\bibfnamefont {B.~H.}\ \bibnamefont
  {Kim}}, \bibinfo {author} {\bibfnamefont {G.}~\bibnamefont {Khaliullin}}, \
  and\ \bibinfo {author} {\bibfnamefont {B.~I.}\ \bibnamefont {Min}},\
  }\href@noop {} {\bibfield  {journal} {\bibinfo  {journal} {Phys. Rev. Lett.}\
  }\textbf {\bibinfo {volume} {109}},\ \bibinfo {pages} {167205} (\bibinfo
  {year} {2012})}\BibitemShut {NoStop}%
\bibitem [{\citenamefont {Hogan}\ \emph {et~al.}(2015)\citenamefont {Hogan},
  \citenamefont {Yamani}, \citenamefont {Walkup}, \citenamefont {Chen},
  \citenamefont {Dally}, \citenamefont {Ward}, \citenamefont {Dean},
  \citenamefont {Hill}, \citenamefont {Islam}, \citenamefont {Madhavan},\ and\
  \citenamefont {Wilson}}]{Hogan}%
  \BibitemOpen
  \bibfield  {author} {\bibinfo {author} {\bibfnamefont {T.}~\bibnamefont
  {Hogan}}, \bibinfo {author} {\bibfnamefont {Z.}~\bibnamefont {Yamani}},
  \bibinfo {author} {\bibfnamefont {D.}~\bibnamefont {Walkup}}, \bibinfo
  {author} {\bibfnamefont {X.}~\bibnamefont {Chen}}, \bibinfo {author}
  {\bibfnamefont {R.}~\bibnamefont {Dally}}, \bibinfo {author} {\bibfnamefont
  {T.~Z.}\ \bibnamefont {Ward}}, \bibinfo {author} {\bibfnamefont {M.~P.~M.}\
  \bibnamefont {Dean}}, \bibinfo {author} {\bibfnamefont {J.}~\bibnamefont
  {Hill}}, \bibinfo {author} {\bibfnamefont {Z.}~\bibnamefont {Islam}},
  \bibinfo {author} {\bibfnamefont {V.}~\bibnamefont {Madhavan}}, \ and\
  \bibinfo {author} {\bibfnamefont {S.~D.}\ \bibnamefont {Wilson}},\
  }\href@noop {} {\bibfield  {journal} {\bibinfo  {journal} {Phys. Rev. Lett.}\
  }\textbf {\bibinfo {volume} {114}},\ \bibinfo {pages} {257203} (\bibinfo
  {year} {2015})}\BibitemShut {NoStop}%
\bibitem [{\citenamefont {Dai}\ \emph {et~al.}(2014)\citenamefont {Dai},
  \citenamefont {Calleja}, \citenamefont {Cao},\ and\ \citenamefont
  {McElroy}}]{Dai2014}%
  \BibitemOpen
  \bibfield  {author} {\bibinfo {author} {\bibfnamefont {J.}~\bibnamefont
  {Dai}}, \bibinfo {author} {\bibfnamefont {E.}~\bibnamefont {Calleja}},
  \bibinfo {author} {\bibfnamefont {G.}~\bibnamefont {Cao}}, \ and\ \bibinfo
  {author} {\bibfnamefont {K.}~\bibnamefont {McElroy}},\ }\href@noop {}
  {\bibfield  {journal} {\bibinfo  {journal} {Phys. Rev. B}\ }\textbf {\bibinfo
  {volume} {90}},\ \bibinfo {pages} {041102(R)} (\bibinfo {year}
  {2014})}\BibitemShut {NoStop}%
\bibitem [{\citenamefont {Nichols}\ \emph {et~al.}(2014)\citenamefont
  {Nichols}, \citenamefont {Bray-Ali}, \citenamefont {Ansary}, \citenamefont
  {Cao},\ and\ \citenamefont {Ng}}]{Nichols2014}%
  \BibitemOpen
  \bibfield  {author} {\bibinfo {author} {\bibfnamefont {J.}~\bibnamefont
  {Nichols}}, \bibinfo {author} {\bibfnamefont {N.}~\bibnamefont {Bray-Ali}},
  \bibinfo {author} {\bibfnamefont {A.}~\bibnamefont {Ansary}}, \bibinfo
  {author} {\bibfnamefont {G.}~\bibnamefont {Cao}}, \ and\ \bibinfo {author}
  {\bibfnamefont {K.-W.}\ \bibnamefont {Ng}},\ }\href
  {http://link.aps.org/doi/10.1103/PhysRevB.89.085125} {\bibfield  {journal}
  {\bibinfo  {journal} {Phys. Rev. B}\ }\textbf {\bibinfo {volume} {89}},\
  \bibinfo {pages} {085125} (\bibinfo {year} {2014})}\BibitemShut {NoStop}%
\bibitem [{\citenamefont {Li}\ \emph {et~al.}(2013)\citenamefont {Li},
  \citenamefont {Cao}, \citenamefont {Okamoto}, \citenamefont {Yi},
  \citenamefont {Lin}, \citenamefont {Sales}, \citenamefont {Yan},
  \citenamefont {Arita}, \citenamefont {Kuneš}, \citenamefont {Kozhevnikov},
  \citenamefont {Eguiluz}, \citenamefont {Imada}, \citenamefont {Gai},
  \citenamefont {Pan},\ and\ \citenamefont {Mandrus}}]{Li2013}%
  \BibitemOpen
  \bibfield  {author} {\bibinfo {author} {\bibfnamefont {Q.}~\bibnamefont
  {Li}}, \bibinfo {author} {\bibfnamefont {G.}~\bibnamefont {Cao}}, \bibinfo
  {author} {\bibfnamefont {S.}~\bibnamefont {Okamoto}}, \bibinfo {author}
  {\bibfnamefont {J.}~\bibnamefont {Yi}}, \bibinfo {author} {\bibfnamefont
  {W.}~\bibnamefont {Lin}}, \bibinfo {author} {\bibfnamefont {B.~C.}\
  \bibnamefont {Sales}}, \bibinfo {author} {\bibfnamefont {J.}~\bibnamefont
  {Yan}}, \bibinfo {author} {\bibfnamefont {R.}~\bibnamefont {Arita}}, \bibinfo
  {author} {\bibfnamefont {J.}~\bibnamefont {Kuneš}}, \bibinfo {author}
  {\bibfnamefont {A.~V.}\ \bibnamefont {Kozhevnikov}}, \bibinfo {author}
  {\bibfnamefont {A.~G.}\ \bibnamefont {Eguiluz}}, \bibinfo {author}
  {\bibfnamefont {M.}~\bibnamefont {Imada}}, \bibinfo {author} {\bibfnamefont
  {Z.}~\bibnamefont {Gai}}, \bibinfo {author} {\bibfnamefont {M.}~\bibnamefont
  {Pan}}, \ and\ \bibinfo {author} {\bibfnamefont {D.~G.}\ \bibnamefont
  {Mandrus}},\ }\href@noop {} {\bibfield  {journal} {\bibinfo  {journal} {Sci.
  Rep.}\ }\textbf {\bibinfo {volume} {3}} (\bibinfo {year} {2013})}\BibitemShut
  {NoStop}%
\bibitem [{\citenamefont {Chen}\ \emph {et~al.}(2015)\citenamefont {Chen},
  \citenamefont {Hogan}, \citenamefont {Walkup}, \citenamefont {Zhou},
  \citenamefont {Pokharel}, \citenamefont {Yao}, \citenamefont {Tian},
  \citenamefont {Ward}, \citenamefont {Zhao}, \citenamefont {Parshall},
  \citenamefont {Opeil}, \citenamefont {Lynn}, \citenamefont {Madhavan},\ and\
  \citenamefont {Wilson}}]{Chen2015}%
  \BibitemOpen
  \bibfield  {author} {\bibinfo {author} {\bibfnamefont {X.}~\bibnamefont
  {Chen}}, \bibinfo {author} {\bibfnamefont {T.}~\bibnamefont {Hogan}},
  \bibinfo {author} {\bibfnamefont {D.}~\bibnamefont {Walkup}}, \bibinfo
  {author} {\bibfnamefont {W.}~\bibnamefont {Zhou}}, \bibinfo {author}
  {\bibfnamefont {M.}~\bibnamefont {Pokharel}}, \bibinfo {author}
  {\bibfnamefont {M.}~\bibnamefont {Yao}}, \bibinfo {author} {\bibfnamefont
  {W.}~\bibnamefont {Tian}}, \bibinfo {author} {\bibfnamefont {T.~Z.}\
  \bibnamefont {Ward}}, \bibinfo {author} {\bibfnamefont {Y.}~\bibnamefont
  {Zhao}}, \bibinfo {author} {\bibfnamefont {D.}~\bibnamefont {Parshall}},
  \bibinfo {author} {\bibfnamefont {C.}~\bibnamefont {Opeil}}, \bibinfo
  {author} {\bibfnamefont {J.~W.}\ \bibnamefont {Lynn}}, \bibinfo {author}
  {\bibfnamefont {V.}~\bibnamefont {Madhavan}}, \ and\ \bibinfo {author}
  {\bibfnamefont {S.~D.}\ \bibnamefont {Wilson}},\ }\href@noop {} {\bibfield
  {journal} {\bibinfo  {journal} {Phys. Rev. B}\ }\textbf {\bibinfo {volume}
  {92}},\ \bibinfo {pages} {075125} (\bibinfo {year} {2015})}\BibitemShut
  {NoStop}%
\bibitem [{\citenamefont {de~la Torre}\ \emph {et~al.}(2015)\citenamefont
  {de~la Torre}, \citenamefont {McKeownWalker}, \citenamefont {Bruno},
  \citenamefont {Ricc{\'{o}}}, \citenamefont {Wang}, \citenamefont {{Gutierrez
  Lezama}}, \citenamefont {Scheerer}, \citenamefont {Giriat}, \citenamefont
  {Jaccard}, \citenamefont {Berthod}, \citenamefont {Kim}, \citenamefont
  {Hoesch}, \citenamefont {Hunter}, \citenamefont {Perry}, \citenamefont
  {Tamai},\ and\ \citenamefont {Baumberger}}]{DelaTorre2015}%
  \BibitemOpen
  \bibfield  {author} {\bibinfo {author} {\bibfnamefont {A.}~\bibnamefont
  {de~la Torre}}, \bibinfo {author} {\bibfnamefont {S.}~\bibnamefont
  {McKeownWalker}}, \bibinfo {author} {\bibfnamefont {F.~Y.}\ \bibnamefont
  {Bruno}}, \bibinfo {author} {\bibfnamefont {S.}~\bibnamefont {Ricc{\'{o}}}},
  \bibinfo {author} {\bibfnamefont {Z.}~\bibnamefont {Wang}}, \bibinfo {author}
  {\bibfnamefont {I.}~\bibnamefont {{Gutierrez Lezama}}}, \bibinfo {author}
  {\bibfnamefont {G.}~\bibnamefont {Scheerer}}, \bibinfo {author}
  {\bibfnamefont {G.}~\bibnamefont {Giriat}}, \bibinfo {author} {\bibfnamefont
  {D.}~\bibnamefont {Jaccard}}, \bibinfo {author} {\bibfnamefont
  {C.}~\bibnamefont {Berthod}}, \bibinfo {author} {\bibfnamefont {T.~K.}\
  \bibnamefont {Kim}}, \bibinfo {author} {\bibfnamefont {M.}~\bibnamefont
  {Hoesch}}, \bibinfo {author} {\bibfnamefont {E.~C.}\ \bibnamefont {Hunter}},
  \bibinfo {author} {\bibfnamefont {R.~S.}\ \bibnamefont {Perry}}, \bibinfo
  {author} {\bibfnamefont {A.}~\bibnamefont {Tamai}}, \ and\ \bibinfo {author}
  {\bibfnamefont {F.}~\bibnamefont {Baumberger}},\ }\href@noop {} {\bibfield
  {journal} {\bibinfo  {journal} {Phys. Rev. Lett.}\ }\textbf {\bibinfo
  {volume} {115}},\ \bibinfo {pages} {176402} (\bibinfo {year}
  {2015})}\BibitemShut {NoStop}%
\bibitem [{\citenamefont {Battisti}\ \emph {et~al.}(2017)\citenamefont
  {Battisti}, \citenamefont {Bastiaans}, \citenamefont {Fedoseev},
  \citenamefont {de~la Torre}, \citenamefont {Iliopoulos}, \citenamefont
  {Tamai}, \citenamefont {Hunter}, \citenamefont {Perry}, \citenamefont
  {Zaanen}, \citenamefont {Baumberger},\ and\ \citenamefont
  {Allan}}]{Battisti2017}%
  \BibitemOpen
  \bibfield  {author} {\bibinfo {author} {\bibfnamefont {I.}~\bibnamefont
  {Battisti}}, \bibinfo {author} {\bibfnamefont {K.~M.}\ \bibnamefont
  {Bastiaans}}, \bibinfo {author} {\bibfnamefont {V.}~\bibnamefont {Fedoseev}},
  \bibinfo {author} {\bibfnamefont {A.}~\bibnamefont {de~la Torre}}, \bibinfo
  {author} {\bibfnamefont {N.}~\bibnamefont {Iliopoulos}}, \bibinfo {author}
  {\bibfnamefont {A.}~\bibnamefont {Tamai}}, \bibinfo {author} {\bibfnamefont
  {E.~C.}\ \bibnamefont {Hunter}}, \bibinfo {author} {\bibfnamefont {R.~S.}\
  \bibnamefont {Perry}}, \bibinfo {author} {\bibfnamefont {J.}~\bibnamefont
  {Zaanen}}, \bibinfo {author} {\bibfnamefont {F.}~\bibnamefont {Baumberger}},
  \ and\ \bibinfo {author} {\bibfnamefont {M.~P.}\ \bibnamefont {Allan}},\
  }\href@noop {} {\bibfield  {journal} {\bibinfo  {journal} {Nat. Phys.}\
  }\textbf {\bibinfo {volume} {13}},\ \bibinfo {pages} {21} (\bibinfo {year}
  {2017})}\BibitemShut {NoStop}%
\bibitem [{\citenamefont {Chikara}\ \emph {et~al.}(2009)\citenamefont
  {Chikara}, \citenamefont {Korneta}, \citenamefont {Crummett}, \citenamefont
  {DeLong}, \citenamefont {Schlottmann},\ and\ \citenamefont
  {Cao}}]{Chikara2009a}%
  \BibitemOpen
  \bibfield  {author} {\bibinfo {author} {\bibfnamefont {S.}~\bibnamefont
  {Chikara}}, \bibinfo {author} {\bibfnamefont {O.}~\bibnamefont {Korneta}},
  \bibinfo {author} {\bibfnamefont {W.~P.}\ \bibnamefont {Crummett}}, \bibinfo
  {author} {\bibfnamefont {L.~E.}\ \bibnamefont {DeLong}}, \bibinfo {author}
  {\bibfnamefont {P.}~\bibnamefont {Schlottmann}}, \ and\ \bibinfo {author}
  {\bibfnamefont {G.}~\bibnamefont {Cao}},\ }\href@noop {} {\bibfield
  {journal} {\bibinfo  {journal} {Phys. Rev. B}\ }\textbf {\bibinfo {volume}
  {80}},\ \bibinfo {pages} {140407(R)} (\bibinfo {year} {2009})}\BibitemShut
  {NoStop}%
\bibitem [{\citenamefont {{COMSOL}}(2015)}]{comsol}%
  \BibitemOpen
  \bibfield  {author} {\bibinfo {author} {\bibnamefont {{COMSOL}}},\
  }\href@noop {} {}\bibinfo {howpublished} {{COMSOL} {M}ultiphysics 5.2}
  (\bibinfo {year} {2015})\BibitemShut {NoStop}%
\bibitem [{\citenamefont {Bardeen}(1961)}]{Bardeen1961}%
  \BibitemOpen
  \bibfield  {author} {\bibinfo {author} {\bibfnamefont {J.}~\bibnamefont
  {Bardeen}},\ }\href@noop {} {\bibfield  {journal} {\bibinfo  {journal} {Phys.
  Rev. Lett.}\ }\textbf {\bibinfo {volume} {6}},\ \bibinfo {pages} {57}
  (\bibinfo {year} {1961})}\BibitemShut {NoStop}%
\bibitem [{\citenamefont {Feenstra}\ \emph {et~al.}(1987)\citenamefont
  {Feenstra}, \citenamefont {Stroscio},\ and\ \citenamefont
  {Fein}}]{Feenstra1987}%
  \BibitemOpen
  \bibfield  {author} {\bibinfo {author} {\bibfnamefont {R.}~\bibnamefont
  {Feenstra}}, \bibinfo {author} {\bibfnamefont {J.~A.}\ \bibnamefont
  {Stroscio}}, \ and\ \bibinfo {author} {\bibfnamefont {A.}~\bibnamefont
  {Fein}},\ }\href@noop {} {\bibfield  {journal} {\bibinfo  {journal} {Surf.
  Sci.}\ }\textbf {\bibinfo {volume} {181}},\ \bibinfo {pages} {295} (\bibinfo
  {year} {1987})}\BibitemShut {NoStop}%
\bibitem [{\citenamefont {Wang}\ \emph {et~al.}(2013)\citenamefont {Wang},
  \citenamefont {Cao}, \citenamefont {Waugh}, \citenamefont {Park},
  \citenamefont {Qi}, \citenamefont {Korneta}, \citenamefont {Cao},\ and\
  \citenamefont {Dessau}}]{Wang2013}%
  \BibitemOpen
  \bibfield  {author} {\bibinfo {author} {\bibfnamefont {Q.}~\bibnamefont
  {Wang}}, \bibinfo {author} {\bibfnamefont {Y.}~\bibnamefont {Cao}}, \bibinfo
  {author} {\bibfnamefont {J.~A.}\ \bibnamefont {Waugh}}, \bibinfo {author}
  {\bibfnamefont {S.~R.}\ \bibnamefont {Park}}, \bibinfo {author}
  {\bibfnamefont {T.~F.}\ \bibnamefont {Qi}}, \bibinfo {author} {\bibfnamefont
  {O.~B.}\ \bibnamefont {Korneta}}, \bibinfo {author} {\bibfnamefont
  {G.}~\bibnamefont {Cao}}, \ and\ \bibinfo {author} {\bibfnamefont {D.~S.}\
  \bibnamefont {Dessau}},\ }\href@noop {} {\bibfield  {journal} {\bibinfo
  {journal} {Phys. Rev. B}\ }\textbf {\bibinfo {volume} {87}},\ \bibinfo
  {pages} {245109} (\bibinfo {year} {2013})}\BibitemShut {NoStop}%
\bibitem [{\citenamefont {Wijnheijmer}\ \emph {et~al.}(2009)\citenamefont
  {Wijnheijmer}, \citenamefont {Garleff}, \citenamefont {Teichmann},
  \citenamefont {Wenderoth}, \citenamefont {Loth}, \citenamefont {Ulbrich},
  \citenamefont {Maksym}, \citenamefont {Roy},\ and\ \citenamefont
  {Koenraad}}]{Wijnheijmer2009}%
  \BibitemOpen
  \bibfield  {author} {\bibinfo {author} {\bibfnamefont {A.~P.}\ \bibnamefont
  {Wijnheijmer}}, \bibinfo {author} {\bibfnamefont {J.~K.}\ \bibnamefont
  {Garleff}}, \bibinfo {author} {\bibfnamefont {K.}~\bibnamefont {Teichmann}},
  \bibinfo {author} {\bibfnamefont {M.}~\bibnamefont {Wenderoth}}, \bibinfo
  {author} {\bibfnamefont {S.}~\bibnamefont {Loth}}, \bibinfo {author}
  {\bibfnamefont {R.~G.}\ \bibnamefont {Ulbrich}}, \bibinfo {author}
  {\bibfnamefont {P.~A.}\ \bibnamefont {Maksym}}, \bibinfo {author}
  {\bibfnamefont {M.}~\bibnamefont {Roy}}, \ and\ \bibinfo {author}
  {\bibfnamefont {P.~M.}\ \bibnamefont {Koenraad}},\ }\href@noop {} {\bibfield
  {journal} {\bibinfo  {journal} {Phys. Rev. Lett.}\ }\textbf {\bibinfo
  {volume} {102}},\ \bibinfo {pages} {166101} (\bibinfo {year}
  {2009})}\BibitemShut {NoStop}%
\bibitem [{\citenamefont {Marczinowski}\ \emph {et~al.}(2008)\citenamefont
  {Marczinowski}, \citenamefont {Wiebe}, \citenamefont {Meier}, \citenamefont
  {Hashimoto},\ and\ \citenamefont {Wiesendanger}}]{Marczinowski2008}%
  \BibitemOpen
  \bibfield  {author} {\bibinfo {author} {\bibfnamefont {F.}~\bibnamefont
  {Marczinowski}}, \bibinfo {author} {\bibfnamefont {J.}~\bibnamefont {Wiebe}},
  \bibinfo {author} {\bibfnamefont {F.}~\bibnamefont {Meier}}, \bibinfo
  {author} {\bibfnamefont {K.}~\bibnamefont {Hashimoto}}, \ and\ \bibinfo
  {author} {\bibfnamefont {R.}~\bibnamefont {Wiesendanger}},\ }\href@noop {}
  {\bibfield  {journal} {\bibinfo  {journal} {Phys. Rev. B}\ }\textbf {\bibinfo
  {volume} {77}},\ \bibinfo {pages} {115318} (\bibinfo {year}
  {2008})}\BibitemShut {NoStop}%
\bibitem [{\citenamefont {Lee}\ and\ \citenamefont {Gupta}(2011)}]{Lee2011}%
  \BibitemOpen
  \bibfield  {author} {\bibinfo {author} {\bibfnamefont {D.-H.}\ \bibnamefont
  {Lee}}\ and\ \bibinfo {author} {\bibfnamefont {J.~A.}\ \bibnamefont
  {Gupta}},\ }\href@noop {} {\bibfield  {journal} {\bibinfo  {journal} {Nano
  Letters}\ }\textbf {\bibinfo {volume} {11}},\ \bibinfo {pages} {2004}
  (\bibinfo {year} {2011})}\BibitemShut {NoStop}%
\bibitem [{\citenamefont {Zheng}\ \emph {et~al.}(2012)\citenamefont {Zheng},
  \citenamefont {Kr{\"{o}}ger},\ and\ \citenamefont {Berndt}}]{Zheng2012}%
  \BibitemOpen
  \bibfield  {author} {\bibinfo {author} {\bibfnamefont {H.}~\bibnamefont
  {Zheng}}, \bibinfo {author} {\bibfnamefont {J.}~\bibnamefont {Kr{\"{o}}ger}},
  \ and\ \bibinfo {author} {\bibfnamefont {R.}~\bibnamefont {Berndt}},\
  }\href@noop {} {\bibfield  {journal} {\bibinfo  {journal} {Phys. Rev. Lett.}\
  }\textbf {\bibinfo {volume} {108}},\ \bibinfo {pages} {076801} (\bibinfo
  {year} {2012})}\BibitemShut {NoStop}%
\bibitem [{\citenamefont {Zheng}\ \emph {et~al.}(2013)\citenamefont {Zheng},
  \citenamefont {Weismann},\ and\ \citenamefont {Berndt}}]{Zheng2013}%
  \BibitemOpen
  \bibfield  {author} {\bibinfo {author} {\bibfnamefont {H.}~\bibnamefont
  {Zheng}}, \bibinfo {author} {\bibfnamefont {A.}~\bibnamefont {Weismann}}, \
  and\ \bibinfo {author} {\bibfnamefont {R.}~\bibnamefont {Berndt}},\
  }\href@noop {} {\bibfield  {journal} {\bibinfo  {journal} {Phys. Rev. Lett.}\
  }\textbf {\bibinfo {volume} {110}},\ \bibinfo {pages} {226101} (\bibinfo
  {year} {2013})}\BibitemShut {NoStop}%
\bibitem [{\citenamefont {Steele}\ \emph {et~al.}(2005)\citenamefont {Steele},
  \citenamefont {Ashoori}, \citenamefont {Pfeiffer},\ and\ \citenamefont
  {West}}]{Steele2005}%
  \BibitemOpen
  \bibfield  {author} {\bibinfo {author} {\bibfnamefont {G.~A.}\ \bibnamefont
  {Steele}}, \bibinfo {author} {\bibfnamefont {R.~C.}\ \bibnamefont {Ashoori}},
  \bibinfo {author} {\bibfnamefont {L.~N.}\ \bibnamefont {Pfeiffer}}, \ and\
  \bibinfo {author} {\bibfnamefont {K.~W.}\ \bibnamefont {West}},\ }\href@noop
  {} {\bibfield  {journal} {\bibinfo  {journal} {Phys. Rev. Lett.}\ }\textbf
  {\bibinfo {volume} {95}},\ \bibinfo {pages} {136804} (\bibinfo {year}
  {2005})}\BibitemShut {NoStop}%
\bibitem [{\citenamefont {Walkup}(2016)}]{Walkup2016}%
  \BibitemOpen
  \bibfield  {author} {\bibinfo {author} {\bibfnamefont {D.~T.}\ \bibnamefont
  {Walkup}},\ }\emph {\bibinfo {title} {PhD Thesis}},\ \href@noop {} {Ph.D.
  thesis} (\bibinfo {year} {2016})\BibitemShut {NoStop}%
\bibitem [{\citenamefont {Ge}\ \emph {et~al.}(2011)\citenamefont {Ge},
  \citenamefont {Qi}, \citenamefont {Korneta}, \citenamefont {De~Long},
  \citenamefont {Schlottmann}, \citenamefont {Crummett},\ and\ \citenamefont
  {Cao}}]{Ge2011}%
  \BibitemOpen
  \bibfield  {author} {\bibinfo {author} {\bibfnamefont {M.}~\bibnamefont
  {Ge}}, \bibinfo {author} {\bibfnamefont {T.~F.}\ \bibnamefont {Qi}}, \bibinfo
  {author} {\bibfnamefont {O.~B.}\ \bibnamefont {Korneta}}, \bibinfo {author}
  {\bibfnamefont {D.~E.}\ \bibnamefont {De~Long}}, \bibinfo {author}
  {\bibfnamefont {P.}~\bibnamefont {Schlottmann}}, \bibinfo {author}
  {\bibfnamefont {W.~P.}\ \bibnamefont {Crummett}}, \ and\ \bibinfo {author}
  {\bibfnamefont {G.}~\bibnamefont {Cao}},\ }\href@noop {} {\bibfield
  {journal} {\bibinfo  {journal} {Phys. Rev. B}\ }\textbf {\bibinfo {volume}
  {84}},\ \bibinfo {pages} {100402(R)} (\bibinfo {year} {2011})}\BibitemShut
  {NoStop}%
\bibitem [{\citenamefont {Jackson}(1999)}]{Jackson1999}%
  \BibitemOpen
  \bibfield  {author} {\bibinfo {author} {\bibfnamefont {J.~D.}\ \bibnamefont
  {Jackson}},\ }in\ \href@noop {} {\emph {\bibinfo {booktitle} {Class.
  Electrodyn.}}}\ (\bibinfo  {publisher} {John Wiley {\&} Sons},\ \bibinfo
  {year} {1999})\ \bibinfo {edition} {third edit}\ ed.,\ pp.\ \bibinfo {pages}
  {154--156}\BibitemShut {NoStop}%
\end{thebibliography}%

\end{document}